\pdfoutput=1

\documentclass[12pt]{iopart}
\usepackage{iopams} 
\usepackage{graphics, graphicx}
\usepackage{hyperref}

\def\ded{\ifletter\bigskip\noindent\ignorespaces\else
    \section*{Dedication}\fi}
    
\newcommand{\ev}{\mbox{$\rm eV$}}
\newcommand{\meV}{\mbox{$\rm meV$}}
\newcommand{\CL}{\mbox{$\rm 90 \% \,C.L.$}}

\newcommand{\eVc}{\mbox{eV$^2$/c$^4$}}

\newcommand{\mnu}{\mbox{$m_\nu$}}

\newcommand{\rhod}{\mbox{$\cal{N}$}}
\newcommand{\rhoz}{\mbox{$n \mbox{(z)}$}}
\newcommand{\rhor}{\mbox{$n (\vec{r})$}}

\newcommand{\pin}{\mbox{$p_{\rm in}$}}
\newcommand{\pout}{\mbox{$p_{\rm out}$}}

\newcommand{\Deltatrue}{\mbox{$\Delta_{\rm{true}}$}} 
\newcommand{\Deltaprec}{\mbox{$\Delta_{\rm{prec}}$}} 
\newcommand{\Deltaacc}{\mbox{$\Delta_{\rm{acc}}$}}

\begin{document}

\title[Monitoring of the operating parameters of the KATRIN WGTS]{Monitoring of the operating parameters of the KATRIN Windowless Gaseous Tritium Source}
%
\author{M. Babutzka$^{3}$, M.~Bahr$^{8}$, J.~Bonn$^{1}$, B.~Bornschein$^{3}$, A.~Dieter$^{2}$, G.~Drexlin$^{1, 2}$, K.~Eitel$^{1}$, S.~Fischer$^{3}$, F.~Gl\"uck$^{2, 5}$, S.~Grohmann$^{4}$, M.~H\"otzel$^{2}$, T.~M.~James$^{6}$, W.~K\"afer$^{1}$\footnote{Corresponding author}, M.~Leber$^{8}$, B.~Monreal$^{8}$,   F.~Priester$^{3}$,  M.~R\"ollig$^{3}$,  M.~Schl\"osser$^{3}$, U.~Schmitt$^{1}$, F.~Sharipov$^{7}$, M.~Steidl$^{1}$\footnotemark[1], M.~Sturm$^{3}$, H.~H.~Telle$^{6}$ and N.~Titov$^{9}$}

\address{$^1$ IKP, Karlsruhe Institute of Technology, P.O. Box 3640, 76021 Karlsruhe, Germany}
\address{$^2$ EKP, Karlsruhe Institute of Technology, P.O. Box 3640, 76021 Karlsruhe, Germany}
\address{$^3$ Tritium Laboratory Karlsruhe,  Karlsruhe Institute of Technology, P.O. Box 3640, 76021 Karlsruhe, Germany} 
\address{$^4$ ITTK,  Karlsruhe Institute of Technology, P.O. Box 3640, 76021 Karlsruhe, Germany}  
\address{$^5$ Institute for Particle and Nuclear Physics, Wigner RCP,\\ H-1525 Budapest, POB 49, Hungary} 
\address{$^6$ Department of Physics, College of Science, Swansea University,\\ 
Singleton Park, Swansea, SA2 8PP, United Kingdom}
\address{$^7$ Departmento de Fisica, Universidade Federal do Parana, Curitiba, Brazil}
\address{$^8$ Department of Physics, University of California, Santa Barbara, US}
\address{$^{9}$ Academy of Sciences of Russia, Institute for Nuclear Research,\\
   $60^{\rm{th}}$ October Anniversary Prospect 7a, 117312 Moscow, Russia}

%

\ead{wolfgang.kaefer@kit.edu, markus.steidl@kit.edu}

\begin{abstract}
The Karlsruhe Tritium Neutrino (KATRIN) experiment will measure the absolute mass scale of neutrinos with a sensitivity of \mnu = 200 meV/c$^2$ by high-precision spectroscopy close to the tritium $\beta$-decay endpoint at 18.6 keV. Its Windowless Gaseous Tritium Source (WGTS) is a $\beta$-decay source of high intensity ($10^{11}$/s) and stability, where high-purity molecular tritium at 30~K is circulated in a closed loop with a yearly throughput of 10~kg. 
To limit systematic effects the column density of the source has to be stabilised at the $10^{-3}$ level. This requires extensive sensor instrumentation and dedicated control and monitoring systems for parameters such as the beam tube temperature, injection pressure, gas composition and others. Here we give an overview of these systems including a dedicated Laser-Raman system as well as several $\beta$-decay activity monitors. We also report on results of the WGTS demonstrator and other large-scale test experiments giving proof-of-principle that all parameters relevant to the systematics can be controlled and monitored on the $10^{-3}$ level or better.  As a result of these works, the WGTS systematics can be controlled within stringent margins, enabling the KATRIN experiment to explore the neutrino mass scale with the design sensitivity.      
\end{abstract}


\submitto{New. J. Phys.} 

\section{Introduction}
With the discovery of massive neutrinos from $\nu$-oscillation experiments about a decade ago (see e.g.~\cite{pdg12} and references therein), one of the most fundamental tasks for the next years will be the determination of the absolute mass scale of neutrinos. 
Neutrino masses can also be obtained from cosmological observations and neutrinoless double beta decay (e.g.~\cite{pdg12}), but these methods strongly depend on the employed model. Model-independent results are provided only by kinematic $\beta$-decay experiments~\cite{Otten-2008}.

The next-generation KArlsruhe TRItium Neutrino experiment (KATRIN) aims at improving the sensitivity in the neutrino mass measurement down to 200~\meV/c$^2$~(\CL)~\cite{TDR}, one order of magnitude lower than upper limits obtained by the Mainz~\cite{mainz-2005} and Troitsk~\cite{troitsk-1999} experiments, the most sensitive direct neutrino mass experiments thus far.
In this kind of experiment, information about the neutrino mass (i.e an upper limit or its value) was and will be derived with a precise measurement of the shape of the tritium $\beta$-spectrum near its endpoint at $E_0=(18571.8\pm1.2)$~eV~\cite{Otten-2008}. The KATRIN experiment will be performed at the Karlsruhe Institute of Technology (KIT). Locating KATRIN at KIT allows the use of the unique expertise of the on-site Tritium Laboratory Karlsruhe (TLK), the only scientific laboratory equipped with a closed tritium loop~\cite{TLK-2005} and licensed to handle the necessary amount of tritium required for the KATRIN experiment.

\begin{figure}[Bbt]
	\centering
		\includegraphics[width=\columnwidth]{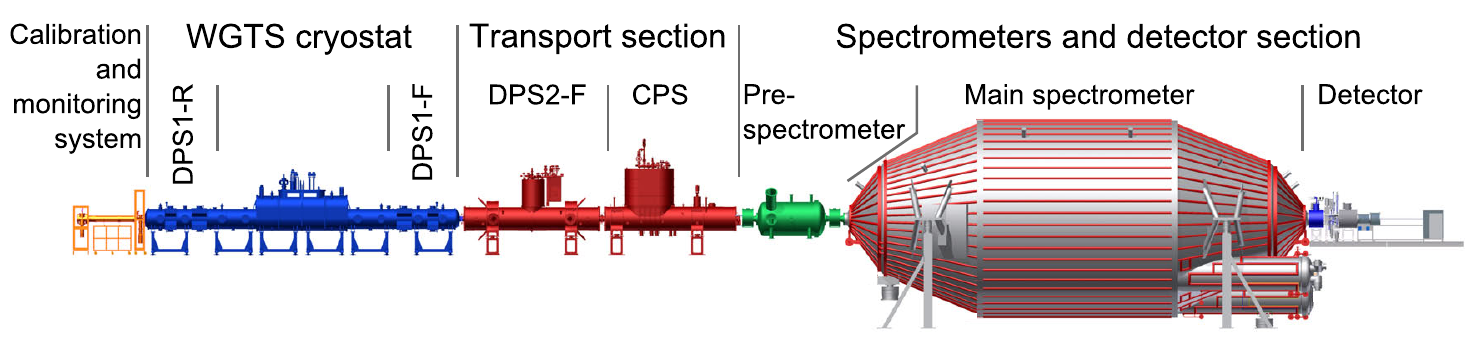}
	\caption{Overview of the KATRIN Experiment. A more detailed description of the system components is found in the main text.}
	\label{fig:katrin-overview}
\end{figure}

The schematic layout of the KATRIN experiment is shown in figure~\ref{fig:katrin-overview}. KATRIN intends to use a Windowless Gaseous Tritium Source (WGTS) with an activity of 10$^{11}$~Bq. Such a source was employed by the LANL experiment~\cite{LANL} and developed further by the Troitsk experiment~\cite{TroitskWGTS}. 
Superconducting solenoids will generate a homogeneous magnetic field of 3.6~T, which adiabatically guides the $\beta$-decay electrons towards the tube ends.

The energy measurement of the electrons is done by the spectrometer and detector section that is attached to the forward side of the WGTS via the transport section. The spectrometers are of MAC-E type~\cite{mac-e,beamson}. The detailed energy analysis is performed by the main spectrometer with an energy resolution of 0.93~\ev. It uses an electrostatic retarding potential to transmit electrons with energies above the chosen retarding energy $qU$ to the detector, while electrons with lower energies are rejected. By changing the retarding potential, KATRIN will measure an integrated tritium $\beta$-spectrum, reaching its design sensitivity of 200~\meV/c$^2$~(\CL) within three full years of neutrino mass data taking, corresponding to about 5 to 6 years of operation.

The continuous precise knowledge or, even better, a high stability of the isotopic composition and the column density of the WGTS is very important for the determination of the neutrino mass. These parameters are associated with the main systematic uncertainties of KATRIN, namely activity fluctuations of the WGTS, energy loss corrections taking into account scattering of $\beta$-electrons in the WGTS, and the final state spectrum~\cite{TDR}. For example, underestimating the energy loss will lead to a negative neutrino mass squared in the analysis, a problem that was encountered in analyses during the nineties of the last century~\cite{Otten-2008, Bornschein-2008}.
Furthermore, the knowledge of these parameters is vital to combine data taken over the extended measurement period of a few years.

Here we present the significant progress made in modelling the gas dynamics inside the WGTS as well as in the construction and commissioning of the actual components for controlling and monitoring the WGTS parameters. We focus on the methodologies and instrumentation needed to control and monitor the column density \rhod\, and tritium purity $\epsilon_T$ as close to real time as possible, i.e. the time to obtain the information is shorter than or equal to the time interval (of the order of minutes) during which the retarding potential is constant before being shifted to the next value in the sequence.  
It should be noted that the different components are well past their design stage and in some cases are already fully operational.

The paper is structured as follows: In section~\ref{sec:Source}, we give an introduction on the WGTS and its important parameters. Our gas dynamics simulation is presented in section~\ref{sec:Modelling} as well as the resulting requirements for the different control and monitoring methods for the WGTS. Their status is described in section~\ref{sec:control-monitoring}. The paper finishes with a presentation of the impact for KATRIN in section~\ref{sec:conclusions}.

\section{The Windowless Gaseous Tritium Source (WGTS)}
~\label{sec:Source}
Key parameters of the WGTS to achieve high statistical sensitivity in the neutrino mass search are the column density $\rhod$, defined as
the number of molecules within a flux tube volume of unit cross section (further discussed in section~\ref{sec:TheColumnDensity}), and the isotopic composition (T${}_2$, DT, HT, D${}_2$, HD, H${}_2$), as the count rate $S$ of the source scales like:
\begin{equation}
~\label{eq:signal}
S=C\cdot \epsilon_T \cdot \rhod. 
\end{equation}
Here $\epsilon_T$ refers to the tritium purity further discussed in section~\ref{sec:IsotopologicalComposition}; the proportionality constant $C$ encompasses various experimental properties (detector efficiency, acceptance, etc.).

It should be noted, that equation~\eref{eq:signal} is not necessarily valid over the full \rhod -range since the observed activity of a source saturates for large column densities. The reason for this is that electrons escape the source only with reduced energy, due to the increasing scattering probabilities, which is enhanced by the increasing cross section towards lower energies; thus these electrons cannot overcome the retarding potential anymore. This can lead to non-linearities in the relation between the measured source activity and column density. However, simulations with full particle tracking of the scattered electrons in the magnetic field down to a threshold of 100~eV show that the non-linearity of equation~\eref{eq:signal} is negligible in an interval $\pm1\%$ around a central value $\rhod=5.0\times10^{17}\mbox{molecules}/\mbox{cm}^2$.  

Both parameters, $\rhod$\,and $\epsilon_T$, also play a significant role for the systematic uncertainty in deriving \mnu \,from the measurement as \rhod \,strongly affects the energy losses of the electrons inside the source by inelastic scattering. Non-T${}_2$ impurities lead to mandatory corrections of the $\beta$-electron spectrum, because of their different recoil energies and different molecular final state distributions.

In the discussion of the requirements for \rhod \, and $\epsilon_T$\, monitoring, we will distinguish between trueness, precision and accuracy (e.g.\cite{ISO}). The trueness \Deltatrue\, is defined as the difference between the mean value of a measured observable and the true value, for example $\Deltatrue (\rhod) = \left\langle \rhod_{{\rm meas}} \right\rangle - \rhod_{{\rm true}}$. The precision \Deltaprec\, corresponds to the square root of the statistical variance of several measured values around their mean value, for instance $\Deltaprec (\rhod)= \sqrt { \left\langle (\rhod_{{\rm meas}} - \left\langle\rhod_{{\rm meas}}\right\rangle)^2 \right\rangle }$. The accuracy \Deltaacc\, is defined as the difference between a single measured value and the true physical value, e.g. $\Deltaacc (\rhod) = \rhod_{{\rm meas}} -\rhod_{{\rm true}}$.

In section~\ref{SubSec_PhysicsofWGTS} the entities $\rhod\,$and $\epsilon_T$ are defined in detail together with their targets in trueness and precision. Those are derived from the KATRIN goal, namely that no single systematic effect gives rise to an uncertainty of $\Delta m^2_\nu<7.5\cdot10^{-3}\eVc$ to the neutrino mass analysis~\cite{TDR}. In section~\ref{sec:physicalwgts} the technical realization of the WGTS is presented with the focus on the required control and monitoring systems to access $\rhod$\, and $\epsilon_T$ experimentally.

\subsection{Key Parameters of the WGTS}
~\label{SubSec_PhysicsofWGTS} 
\begin{figure}[bt]
	\includegraphics[width=1.0\columnwidth]{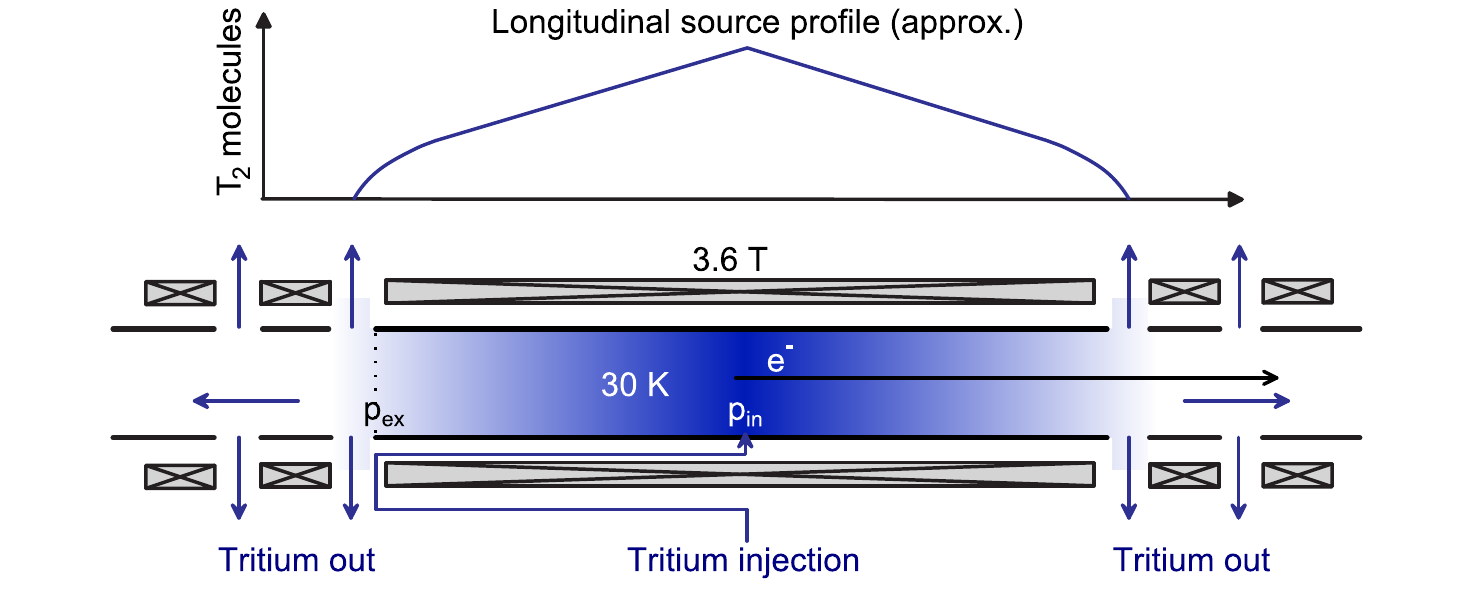}
	\caption[Principle of KATRIN's Windowless Gaseous Tritium Source.]{Principle of the Windowless Gaseous Tritium Source (WGTS). The density profile inside the beam tube is kept constant by continuously injecting tritium gas in the middle and pumping out at both ends.}
	\label{fig:wgts_setup}
\end{figure}

From a physicist's point of view, the WGTS is a column of tritium gas inside an open cylinder with a diameter of $D=90$~mm and a length of $L=10$~m inside a homogeneous magnetic field of $B=3.6$~T generated by superconducting solenoid magnets (see figure~\ref{fig:wgts_setup}). The gas consists of different species with mole fractions $c_i$. 
It is injected near the middle with an adjustable pressure \pin of about $10^{-3}\mbox{mbar}$, and pumped out at both ends with constant pumping speed with an outlet pressure $\pout \approx 0.05 \cdot \pin$. Apart from the geometry and this pressure gradient, the flow for a gas with viscosity $\eta$~ is determined by the beam tube temperature, which has a certain profile $T_{BT}(\phi,z)$ with a mean value in the 30~K regime. Here, $z$ denotes the cylinder axis with $z=0$ being the middle in the source, $\phi$ the azimuthal angle. \\
These parameters define the distribution function $f_i(\vec{r}, \vec{v})$ describing the number of molecules of species $i$, which can be found in the phase space volume $[\vec{r}, \vec{r} + {\rm d}^3\vec{r}], [\vec{v}, \vec{v}+{\rm d}^3\vec{v} ]$.  From calculations of $f_i(\vec{r}, \vec{v})$, the influence of the above parameters on the column density \rhod \, can be investigated. Details on this are given in section~\ref{subsec:Modelling_of_Gasdynamics}.

The full computation of the distribution function $f_i(\vec{r}, \vec{v})$ is also required for investigations of plasma and ion physics effects (e.g. calculations of space charge effects~\cite{plasma}) or for investigating the interactions of low-energy secondary or shake-off electrons due to the short path length of such electrons.
Also, the velocity distribution of the mother molecules is determined by the aforementioned $f_i(\vec{r}, \vec{v})$. Thus one may apply thermal Doppler-broadening corrections to the $\beta$-electrons associated with the thermal diffusion and position-dependent gas flow speeds of the decaying tritium molecules (see section~\ref{subsec:Modelling_of_Gasdynamics}). Such corrections are essential for the exact description of the $\beta$-electron spectrum.

\subsubsection{Column density}
\label{sec:TheColumnDensity}
The column density \rhod \, is defined as
\begin{equation}
~\label{eq:rhod}
	\rhod (\pin, \pout, T_{BT}(\phi,z))= \sum_i \int\limits_{-L/2}^{+L/2} n_i(\rho,\phi,z, \pin, \pout, T_{BT}(\phi,z)) {\rm d}z
\end{equation}
with $n_i(\rho,\phi,z, \pin, \pout, T_{BT}(\phi,z))$ being the integral of $f_i(\vec{r}, \vec{\mbox{v}})$ over the velocity phase space in cylindrical polar coordinates $\left( \rho, \phi, z \right)$ for a given set of experimental boundary conditions. In fact \rhod \, is of course a function of $(\rho
,\phi)$, but due to the homogeneity in $(\rho,\phi)$ (see section~\ref{sec:Implications}) this dependence is omitted in the notation.

At first glance, the knowledge of the true value $\rhod_{{\rm true}}$ seems of no particular interest as the signal strength and therefore \rhod \, (see equation~\eref{eq:signal}) is an unrestricted parameter in the fitting algorithms of the neutrino mass analysis as long as activity fluctuations during one scan of the $\beta$-spectrum are monitored and taken into account. However, the neutrino mass analysis depends on the accurate description of inelastic scattering of the signal electrons by the gas molecules inside the source ($\sigma_{{\rm inel}}(18.6~{\rm keV})~=~3.40(7)\cdot10^{-18}~{\rm cm}^2$)~\cite{Aseev2000}.
Under nominal conditions, 41.3\% of all electrons escape the source without inelastic interaction, 29.3\% are scattered once and 16.7\% are scattered twice. Hereby, the reference column density of the WGTS of $\rhod=5.0\times10^{17}\mbox{molecules}/\mbox{cm}^2$ and the nominal $\vec{B}$-field configuration of KATRIN, which accepts $\beta$-electrons with starting angles with respect to the $\vec{B}$-field of up to $51^{\circ}$ for transmission through the KATRIN beam line, is assumed.
In principle, these scattering probabilities and associated energy losses will be measured directly in a dedicated calibration measurement~\cite{TDR}. However, such a measurement is time-consuming and thus cannot be applied in parallel to the neutrino mass search. Hence, monitoring the amount of energy losses due to scattering is required. This is achieved by continuously measuring \rhod \, with an experimental precision $\Deltaprec (\rhod)$.

According to neutrino mass sensitivity simulations, the necessary precision $\Deltaprec (\rhod)/\rhod$\, to keep the corresponding systematic error below $\Delta m^2_\nu<7.5\cdot10^{-3}\eVc$~\cite{TDR}
depends on details of the neutrino mass analysis, in particular the width of the analyzed energy window below the endpoint. For reference, we use the most demanding value from~\cite{TDR}, namely $\Deltaprec (\rhod) / \rhod <0.1\%$.

As can be seen from equation~\eref{eq:rhod}, changes in \rhod\, are due to shifts or fluctuations of the experimental conditions of inlet and outlet pressures $\Delta\pin, \Delta\pout$\, and temperature changes $\Delta T(r,\phi)$. The latter can either be localized or affect the whole beam tube.

\subsubsection{Isotopologue composition}
\label{sec:IsotopologicalComposition}
  
The tritium purity $\epsilon_T$ is defined as the ratio of the number of tritium atoms to the total sum of atoms in the WGTS. A high tritium purity is primarily necessary to maximize the signal rate $S$ according to equation~\eref{eq:signal}.
It is planned to maintain $\epsilon_T \geq 0.95$ over the complete measurement period. This will be achieved using molecular tritium (T$_2$) as the main gas constituent with a high mole fraction $c(\rm{T_2})\geq 0.9$, a small admixture of DT ($c({\rm{DT}}) < 0.1$) and only trace amounts of HT, D${}_2$ and H${}_2$. The tritiated hydrogen isotopologues T${}_2$, DT and HT differ not only in their mass and therefore in their respective recoil energies, but also in the rotational and vibrational final state distribution of their daughter molecules following tritium decay. Both effects are taken into account in the modelling of the WGTS. The corresponding final state distribution for T${}_2$ for the first four initial rotational states and DT/HT for the initial states J = 0 and 1 have been calculated in~\cite{Doss2006}.  

The systematic uncertainty for $m_{\nu}^2$ that is introduced by the theoretical description of the different final states is estimated to be $6\cdot10^{-3}\eVc$~\cite{TDR}. In the neutrino mass analysis, the final state distributions of the individual isotopologues will be weighted with the measured mole fraction of the individual isotopologue. The accuracy of the gas composition measurement therefore has to be converted into a systematic uncertainty $\Deltaacc({\rm gas})$ for $m_{\nu}^2$ which has to be combined with the uncertainty of the final state description, i.e. $(\Delta_{{\rm acc}}(m_{\nu}^2))^2=(6\cdot10^{-3}\eVc)^2 + (\Deltaacc({\rm gas}))^2$. The uncertainty $\Deltaacc({\rm gas})$ should be below a certain limit ($1\cdot10^{-3}\eVc$) in order to avoid a significant inflation of the combined uncertainty even in case of an improvement of the theoretical description of the final state distributions. Simulations~\cite{Schloesser_proceeding} show that an accuracy of $\Deltaacc(\epsilon_T)\approx 1\%$ is required, which can also be derived from general considerations~\cite{Otten-2008}. This is a rather moderate requirement due to the envisaged high ${\rm T_2}$ mole fraction ($\epsilon_T \geq 0.95$).

Regardless of this, there is compelling reason to measure $\epsilon_T$ with at least a magnitude better precision. Namely, it opens up a complementary approach for monitoring $\Delta (\rhod) \,$ which is independent of hydrodynamical calculations of $f_i$. According to equation~\eref{eq:signal}, monitoring \rhod\ is equivalent to monitoring both $\epsilon_T$ and the activity of the WGTS. For this reason the WGTS instrumentation is extended by different activity detectors and a Laser Raman spectroscopy system (LARA) to measure $\epsilon_T$ with a relative precision of $\Deltaprec (\epsilon_T) /\epsilon_T<0.1\% $.

\subsection{The WGTS cryostat} \label{sec:physicalwgts}
Achieving the proposed tritium purity and activity in the WGTS constitutes a rather challenging task. Specifically, the column density inside the source beam tube has to be kept constant to a level of at least 0.1~\%. Accordingly, a variety of parameters have to be stabilized; these are:
\begin{itemize}
\item[-] the injection rate,
\item[-] the pumping speed in forward and rear direction,
\item[-] the gas composition,
\item[-] and the beam tube temperature.
\end{itemize}
Fulfilling these requirements results in a complex cryostat-system, the so-called WGTS cryostat~\cite{grohmann_design}. Its cryogenic system consists of 13 fluid circuits operated with 6 cryogenic fluids. The cryostat requires a measurement and control system with more than 500 sensors. 
Besides the actual WGTS, the cryostat incorporates the DPS1\hbox{-}F and DPS1\hbox{-}R (the first stages of the differential pumping sections in the front and rear direction, respectively). Also included are the first seven superconducting magnets guiding the electrons adiabatically from the source beam tube through the transport section to the spectrometers.

The DPS1-R will be connected to the Calibration and Monitoring System (CMS) via the so-called rear-wall. The rear-wall separates the WGTS beam tube from the CMS and defines the electrical potential of the tritium plasma inside the WGTS. Furthermore, it may be used to measure the $\beta$-electron current directly (see section~\ref{sec:Faraday}), provided the current measurement can be shielded from low energetic secondary electrons and ions. In addition, it acts as an X-Ray converter which allows activity measurements of the source via Beta-Induced X-Ray Spectroscopy (BIXS, see section~\ref{sec:trirex}). Furthermore, the CMS houses an e-gun for calibration measurements (see section~\ref{sec:e-gun}). In the forward direction, the DPS1-F is connected to the DPS2-F, a further differential pumping section. Its role is to reduce the tritium flow by at least another four orders of magnitude~\cite{Lukic20121126} and to host an FT-ICR (Fourier Transform  Ion cyclotron resonance) mass spectrometer for ion physics~\cite{fticr}.

The basic concept of keeping the injection rate into the WGTS constant is based on injecting tritium gas from a pressure controlled buffer vessel over a 5~m long capillary with constant conductance and pumping the gas at both ends of the beam tube.  The details of the complex tritium gas handling system, the so-called Inner Loop, are discussed in section~\ref{sec:loops}.

The Inner Loop is supplied with high purity tritium by the Outer Loop System, which is covered by the existing infrastructure of TLK~\cite{TLK-2005}. The isotopic purity in the Inner Loop System is monitored in-line with a Laser Raman Spectroscopy setup (LARA) discussed in detail in section~\ref{sec:LARA}.

\begin{figure} [tbp]
	\includegraphics[width=\columnwidth]{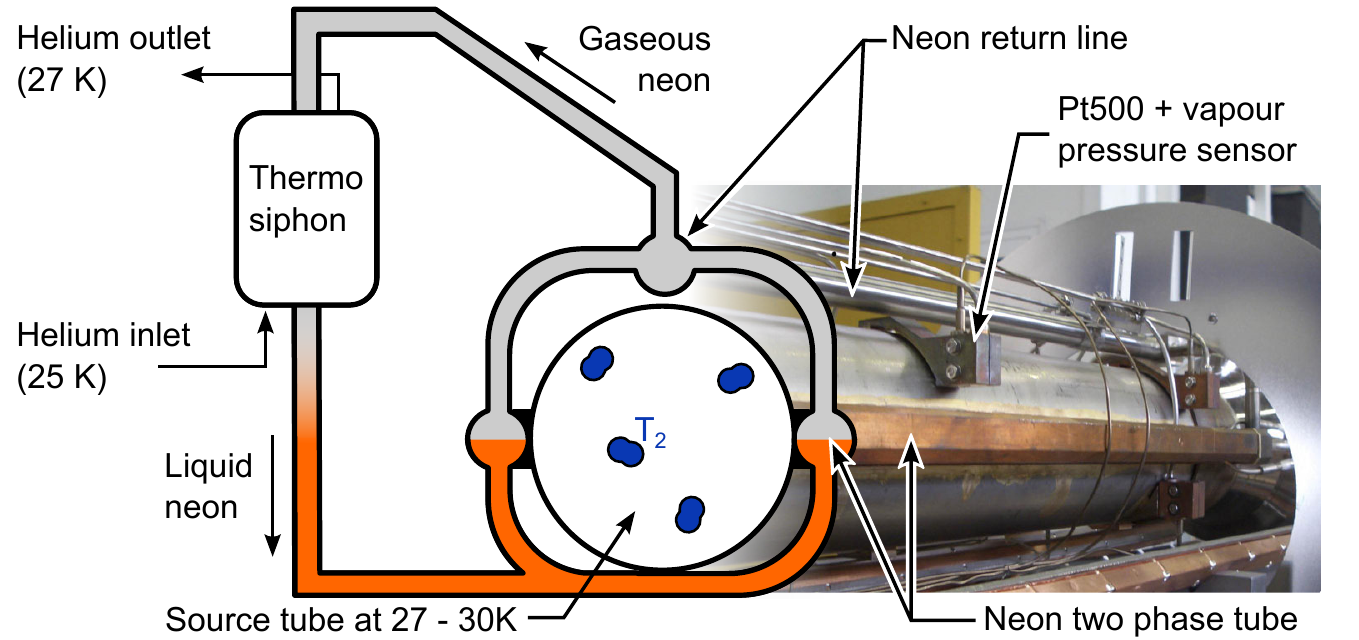}
	\caption{The cooling concept of the WGTS beam tube is based on a thermo-siphon with two 16~mm evaporator tubes brazed onto the beam tube, which are partially filled with neon at its boiling point (30~K). 
	Convection in the thermo-siphon will be controlled by heater wires inside the evaporator tubes. The vapour flows to the evaporator ends, and back through a common return line to the condenser. The thermo-siphon is cooled with liquid helium.}
	\label{fig:thermosiphon}
\end{figure}

The set-point for the WGTS temperature is about 30~K. This constitutes a near optimum: at higher temperatures, the gas flow rate and  the throughput increases. Thus a larger T$_{2}$ inventory would be required to reach the same column density; also the Doppler broadening effect increases. At lower temperatures, clustering and condensation of the hydrogen isotopologues would be induced which would be equally undesirable. 

The concept of providing the cooling for the 10~m long source tube is summarized in figure~\ref{fig:thermosiphon}. As shown there, it is based on a two phase neon thermo-siphon.
Two copper tubes are brazed over the whole length on the outside of the source beam tube and thus determine the temperature on both sides. The neon gas flows to a condenser in a closed cycle; there it is liquefied again, being cooled with gaseous helium at $\sim$25~K. The measured fluctuations of the helium cryogenic circuit used to cool the condenser are of the order of 300~mK~per~hour~\cite{grohmann_stability_analyses}. These fluctuations exceed the temperature stability requirements derived in the next section. Therefore, these fluctuations need to be damped by the coupling between the helium circuit and the two phase neon system, therefore the condenser has a large heat capacity. Since such a cooling concept has never been applied on a large component like the WGTS, a full scale test experiment (the WGTS demonstrator) has been performed in 2011. This demonstrator consists of original components of the final WGTS assembly, in particular the beam tube and its instrumentation. The main difference to the final WGTS cryostat is that all superconducting magnet components are replaced by a cold mass to simulate their thermal properties during the demonstrator tests. Furthermore, the WGTS demonstrator was operated without external potential while the WGTS in its final configuration can be operated on a variable voltage of up to 1000~V to define the potential difference between WGTS and main spectrometer. Selected results of the demonstrator experiment are briefly summarized in section~\ref{sec:BTTTemperatures} and will be discussed in more detail in a future publication~\cite{grohmann_demonstrator}. 
 
The aim of the demonstrator tests is to understand the thermal behaviour of the system and to verify that the only non-negligible influx of heat into the beam tube originates from the pump ports at both ends of the WGTS tube, as predicted and shown in~\cite{grohmann_stability_analyses}. This heat influx leads to an azimuthal temperature profile of the beam tube that can be measured by temperature sensors distributed along the tube and on several azimuthal positions. The measured temperature distribution can then be used in the gas dynamics modelling for the WGTS.


\section{Modelling of the WGTS gas dynamics}
\label{sec:Modelling}

The first part of this chapter, section~\ref{subsec:Modelling_of_Gasdynamics}, focuses on the current state of the gas dynamics simulation developed specifically for the WGTS to compute the density profile and column density, as well as the velocity profile of molecules in the source. From these, one can determine the influence of experimental parameters on the column density and thus ultimately on the neutrino mass sensitivity. In addition, one can deduce the specification for the monitoring of the crucial WGTS parameters presented in section~\ref{sec:Implications}.

\subsection{The gas dynamical model}
\label{subsec:Modelling_of_Gasdynamics}

In order to determine the influence of the thermodynamical properties of the WGTS on the column density \rhod \, as defined by equation~\eref{eq:rhod}, a detailed gas dynamical simulation of the WGTS has been developed, which allows one to compute the density distribution \rhor \, and consequently \rhod. The velocity distribution $f(\rho, \phi, z, \vec{v})$ is also of interest, since the resulting electron energy is Doppler-shifted due to the motion of the tritium molecules. 

One can characterize the flow regime in the WGTS beam tube with the so-called rarefaction parameter~\cite{SHARIPOV98}
\begin{equation}
\delta = \frac{Rp}{\eta\ v_m}, \qquad {\rm with} \   v_m = \sqrt{\frac{ 2 k_B T}{m}}.
\label{eq:delta_orig}
\end{equation}
Here $R$ is the tube radius, $p$ the pressure, $\eta$ the viscosity and $v_m$ is the most probable speed for molecules with mass $m$ and temperature $T$.
Dependent on the flow regime one can distinguish three cases:
\begin{itemize}
	\item[-] The hydrodynamic regime with $\delta \gg 1$, where the equations of continuum mechanics are valid.
	\item[-] The free molecular flow regime ($\delta \ll 1$), where intermolecular collisions can be neglected and the molecules move independently of each other.
	\item[-] The transition regime, where the full Boltzmann equation needs to be solved, using a detailed model of intermolecular collisions. Since for this the intrinsic nature of the intermolecular interactions is important, this is the most difficult scenario.
\end{itemize}
The rarefaction parameter varies from $\delta\approx 20$ at the injection in the WGTS beam tube to $\delta \lesssim 10^{-1}$ at the beginning of the differential pumping chambers, with a significant part of the beam tube exhibiting values within the intermediate regime. One can use the phenomenological intermediate conductance formula of Knudsen~\cite{knudsen} to compute the gas distribution approximately inside the WGTS, delivering an easily computable formula for the pressure distribution in~\cite{ference}. To be more precise and to be able to include detailed boundary conditions, in principle the full Boltzmann equation needs to be solved. 

The length of the beam tube $L$ is much larger than its radius $R$. Thus, to a good approximation, the radial dimensions can be neglected (\rhor = \rhoz) as it will be shown in subsection \ref{sec:Implications}. The problem is reduced to only a single (the longitudinal) spatial dimension, which simplifies the analytical description and tremendously reduces the necessary CPU time required for the calculation. 

For this one-demensional case, the method outlined in~\cite{SHARIPOV97} can be applied: It relies on the fact that the mass flow rate
\begin{equation}
\dot{M} = \frac{ \pi R^3} {v_m} \left[ -G_P(\delta) \frac{ {\rm d}p}{{\rm d} z} + G_T(\delta) \frac{p(z)}{T(z)} \frac{ {\rm d}T}{{\rm d}z} \right],
\label{eq:massflowdiff}
\end{equation}
driven by small gradients of pressure $p$ and temperature $T$, has to remain constant along the tube. 
The Poiseuille coefficient $G_P$ and the thermal creep coefficient $G_T$, which depend only on the rarefaction parameter $\delta$, are available in~\cite{SHARIPOV98}. 
The coefficients were obtained by linearising the Boltzmann equation around a Maxwell-Boltzmann distribution, employing the S-model~\cite{Shakov} for intermolecular collisions and using the small pressure and temperature gradients as perturbation parameters.

\begin{figure}[tb]
\includegraphics[width=\columnwidth]{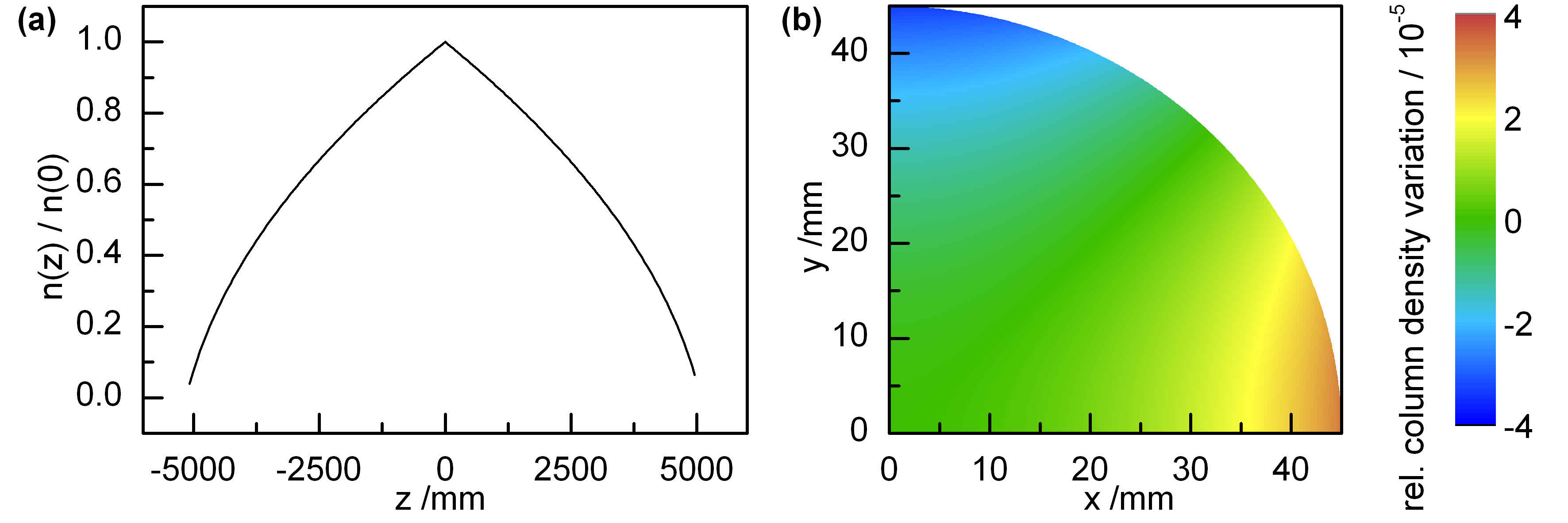}
\caption{ \textsl{(a)}: Calculated relative density profile $n(z)/n(0)$ along the WGTS beam based on an expected source temperature distribution, an injection pressure of $p_{\rm{in}} = 3.368\cdot 10^{-3} \rm{ mbar}$ at $z=0$ and exit pressures of $p_{\rm{out}}= 0.05\, p_{\rm{in}}$ at $z=5\,\rm{m}$. \textsl{(b)}: The expected relative deviations from the average column density $\left( \rhod(x,y) - \rhod_{\rm{ave}}\right) / \rhod_{\rm{ave}} $ for a quarter of the WGTS cross section and an assumed azimuthal temperature variation. 
}
\label{fig:density}%
\end{figure}

With these coefficients, and incorporating the boundary conditions $T(z)$ and \pin\, (cf. sections~\ref{sec:loops} and~\ref{sec:BTTTemperatures}), provided by the continuous monitoring systems, equation~\eref{eq:massflowdiff} can be solved numerically. As a result one can determine the pressure $p = p(z)$, and density $n(z) = p(z)/k_B T(z)$ profiles along the tube.
The resulting density profile obtained for typical values of these boundary conditions is shown in the left part of figure~\ref{fig:density}.
 
However, the one-dimensional model cannot fully describe the gas flow in the more complex geometries of the injection and pump port region. Furthermore, the injection and exit pressures cannot be measured directly, but have to be inferred from pressure sensors in the Inner Loop. Therefore, the development of more detailed three-dimensional simulations of these regions is currently in progress. As a first step, a radial density distribution caused by an azimuthal temperature variation was computed for different positions along the beam tube. Weighting these cross sections with their tritium content allows the computation of the expected radial column density fluctuations shown in the right part of figure~\ref{fig:density}. 

It should be stressed that the gas dynamics simulation will not be used to quantitatively determine the absolute value of \rhod. Although it is, in principle, possible to compute the column density for a given set of boundary conditions $T(z), \pin$ and $\pout$, the simulation has only a trueness of about 5\%. The trueness is limited by two facts: Firstly, important fluid dynamical properties like the viscosity of tritium and the accomodation coefficient between tritium and the beam tube walls are not known accurately enough. Secondly, intrinsic uncertainties of the simulation, in particular the modelling of the intermolecular collisions and using the linearized Boltzmann equation to compute the coefficients $G_P$ and $G_T$ in equation~\eref{eq:massflowdiff}, also cause systematic uncertainties within the simulation. Solving the exact Boltzmann equation numerically has only been achieved in very few and simple cases; the complex WGTS case would require excessive computational effort and is thus not pursued. 
Instead, the gas dynamics simulation is used to determine the influence of a certain (experimental) parameter on \rhod. Once this influence is known, the task of monitoring \rhod \, is translated into continuously measuring this parameter with the required precision. 
In particular the existing one-dimensional simulation is sufficiently fast to be suitable as a nearly real-time analysis tool for the column density based on the actual real-time monitoring of the beam tube temperature and injection pressure.

\subsection{Results of the gas dynamics simulation}
\label{sec:Implications}
Of particular interest for the monitoring requirements of the WGTS 
is the influence of an experimental parameter $X = T_0, \pin, \pout$ on the column density defined in equation~\eref{eq:rhod}, which governs the energy loss of $\beta$-electrons in the source: 
\begin{equation}
\frac{\Delta \rhod}{\rhod} = \alpha_X \frac{\Delta X}{ X}.
\label{eq:prop}
\end{equation}
This influence has been determined with the gas dynamics simulation in~\cite{TDR} and is summarized here. The coefficients $\alpha_X$ given in table~\ref{tbl:propconst} allow for determining the stability requirements for the corresponding physical quantity.
\begin{table}[tbp]%
\centering
\caption{Proportionality constants between different experimental parameters and the column density~\cite{TDR}.}
\begin{tabular}{l l}
\hline
parameter & $\alpha_X$ \\
\hline
temperature & $\alpha_T = -1.2$ \\
injection pressure & $\alpha_{p_{\rm{in}}} = 1.1$  \\
exit pressure & $\alpha_{p_{\rm{out}}} = 0.03$ \\
\hline
\end{tabular}
\label{tbl:propconst}
\end{table}

With the current version of the gas dynamics simulation, it has become possible to include the influence of a temperature profile on \rhor \, and \rhod \, too, as determined by the instrumentation on the WGTS beam tube (see section ~\ref{sec:BTTTemperatures}). Now, it is also possible to compute the full distribution function $f(z, \vec{v})$ instead of only the density distribution. It turns out that the velocity distribution of the tritium molecules is well described analytically using a so-called local Maxwell-Boltzmann distribution 
\begin{equation}
f(\rho, \phi, z, \vec{v}) = \frac{n(z)}{ \left(\sqrt{\pi} v_m \right)^3} \exp{\left(-  \frac{v_\rho^2 + v_\phi^2  + \left(v_z - U_z(\rho, \phi, z) \right)^2}{v_m^2}\right) }.
\label{eq:velocityprof}
\end{equation}
	 The bulk velocity $U_z(\rho, \phi, z)$ reflects the fact that the gas flows from the injection in the middle of the beam tube to both ends. $U_z(\rho, \phi, z)$ is of the order of $10 \,  \rm{ m/s}$ for most of the beam tube length, but increases to about $50\, \rm{m/s}$ at the beam tube exits. This is much smaller than the most probable thermal speed $v_m = \sqrt{ 2k_B T/m_X} \approx 288\, {\rm m/s}$, where $m_X$ is the molecular mass of the respective isotopologue. According to the Doppler effect, the motion of the decaying molecule changes the $\beta$-electron energy as
\begin{equation}
\Delta E = E_{\rm{LAB}} - E_{\rm{CMS}} = \frac{1}{2} m_e \left[\left( \vec{v_{\rm{T_2}}} + \vec{v_e} \right)^2 - \vec{v_e}^2 \right]. 
\label{eq:Doppler}
\end{equation}
 For example, the additional energy change $\Delta E$ for an 18.6~keV $\beta$-electron emitted by a tritium molecule with velocity $v_m$ corresponds to $\Delta E = \cos \theta_{Te} \cdot 130\,\rm{meV}$, where $\theta_{Te}$ is the angle between momenta of the mother molecule and the $\beta$-electron. Since this is of the order of the KATRIN neutrino mass sensitivity, it needs to be taken into account in the $\beta$-spectrum calculation. 

Several additional investigations have been carried out to study specific effects that potentially influence the column density:
\begin{itemize}
	\item A detailed simulation of the injection region based on~\cite{SHARIPOV04} has been performed to optimize the design of the injection chamber; a design with many small holes was shown to be advantageous to avoid turbulence. The final design of the injection chamber features 415~holes with a diameter of 2~mm.
	\item The thermal velocity of the different hydrogen isotopologues differs slightly due to their mass difference $\left(v_m^{\rm{DT}} = \sqrt{m_{\rm{T_2}}/m_{\rm{DT}}} v_m^{{\rm{T_2}}} \approx 1.1 v_m^{{\rm{T_2}}}\right)$. This results in separation phenomena and consequently a dependence of the gas composition on the position. This was studied in dedicated simulations based on~\cite{SHARIPOV11}. In the standard operational mode of the WGTS, these effects are negligible (the relative change is below $10^{-4}$). However, in a special operational mode of the WGTS, the so-called Krypton Mode, this effect becomes significant. In the Krypton Mode, a small fraction of ${}^{83}\rm{Kr}$ is added to the gas mixture for calibration purposes. This implies that the WGTS has to be operated at 120~K since krypton freezes at lower temperatures. In this case, the larger temperature in combination with the larger mass of krypton increases de-mixture effects significantly, but this is not relevant for the work described here.
	\item The influence of an azimuthal temperature variation was investigated and can be calculated for inhomogeneities up to a few K by solving the linearized Boltzmann equation using the discrete velocity method outlined e.g. in~\cite{SHARIPOV98, GRAURSHARIPOV08, Sharipov09}. However, as shown in figure~\ref{fig:density}, this effect is negligible for the expected radial temperature gradients since the corresponding column density fluctuations are below $10^{-4}$.
\end{itemize}

Overall, according to these simulations, the influence of the corresponding effects on \rhor\, and therefore \rhod\, is of the order of $\Delta \rhod / \rhod \lesssim 10^{-4}$; this is safely below the critical value for an unaccounted shift of $ \Delta \rhod_{crit} / \rhod = 0.1\%$~\cite{TDR}. Consequently, monitoring $T(z), p_{\rm{in}}, p_{\rm{out}}$ is sufficient.
 
 The effect on the shape of the tritium $\beta$-spectrum by tritium decays occurring in the pump ports or in the differential pumping section (DPS1-F, DPS1-R) has not yet been investigated in detail. In~\cite{TDR}, it has been noted that effects from decays outside the central beam tube should in principle not be significant due to their small fraction $(p_{\rm{out}}<0.05p_{\rm{in}})$. To properly account decays from these regions, which are partially operated at 80 K and which have inhomogeneous magnetic fields, a simulation with full 3-dimensional geometry has to be done in the future. 
 
The resulting stability requirements for the monitoring systems are summarized in table~\ref{Tbl:monitoring}. In addition to the precision requirements obtained with the values from table~\ref{tbl:propconst}, the absolute value of the beam tube temperature is also of interest; it influences the integrated $\beta$-spectrum through the Doppler effect. 
However, the impact of the Doppler effect is relatively small due to the low source temperature. In recent investigations, we have shown that the resulting accuracy requirement is only about 500~mK. 

For completeness, we briefly address the requirements for gas composition and activity monitoring. As outlined in section~\ref{sec:TheColumnDensity}, monitoring these two quantities constitutes a complementary approach to the monitoring of $p_{\rm{in}}, p_{\rm{out}}$ and $T(z)$, since column density fluctuations can also be obtained by measuring both the gas composition and the source activity using equation~\eref{eq:signal}. Furthermore, as a consequence of the different final state distributions of T${}_{2}$, DT and HT, now also the trueness (and not only the precision) of the gas composition measurement is of relevance. Fortunately, the resulting accuracy requirement is much less stringent than for the precision; in recent studies, we have shown that in case of the envisaged high tritium purity $\epsilon_T\geq0.95$, a $1\%$ accuracy for $\epsilon_T$ can be reached with a trueness of the LARA system of only about $20\%$~\cite{Schloesser_proceeding}.

\begin{table}[tb]%
\caption{Reference values and requirements for the stability of WGTS parameters, partially from~\cite{TDR}.}
\begin{tabular}{l l l l }
\hline
parameter & reference value & req. relative stability & req. trueness\\
\hline
temperature & $T_0=(27-30) \rm{ K}$ & $10^{-3}$ & 0.5\rm{ K} \\
injection pressure & $ p_{\rm{in}} = 3.368 \cdot 10^{-3}\rm{ mbar}$ & $10^{-3}$ & \\
exit pressure & $p_{\rm{out}} \approx (0.03-0.05) p_{\rm{in}}$ & $0.03$ & \\
purity & $ \epsilon_{\rm{T}} >0.95 $& $10^{-3} $ & $0.1$  \\
activity & $A= 10^{11} \rm{ Bq}$& $10^{-3} $ & \\
\hline
\end{tabular}
\label{Tbl:monitoring}
\end{table}


\section{Control and monitoring} \label{sec:control-monitoring}

In table~\ref{Tbl:monitoring} above, the actual requirements for the control and monitoring of the crucial source parameters are listed. These requirements can only be met if the gas inlet and outlet rates, the beam tube temperature and the gas composition are stable. 
In the Technical Design Report of KATRIN~\cite{TDR} only rough concepts 
were sketched. Now that research and development has been performed on all relevant control and monitoring devices, 
it has been demonstrated that these requirements can be fully met. 

\begin{figure}[Tb]
	\centering
		\includegraphics[width=\columnwidth]{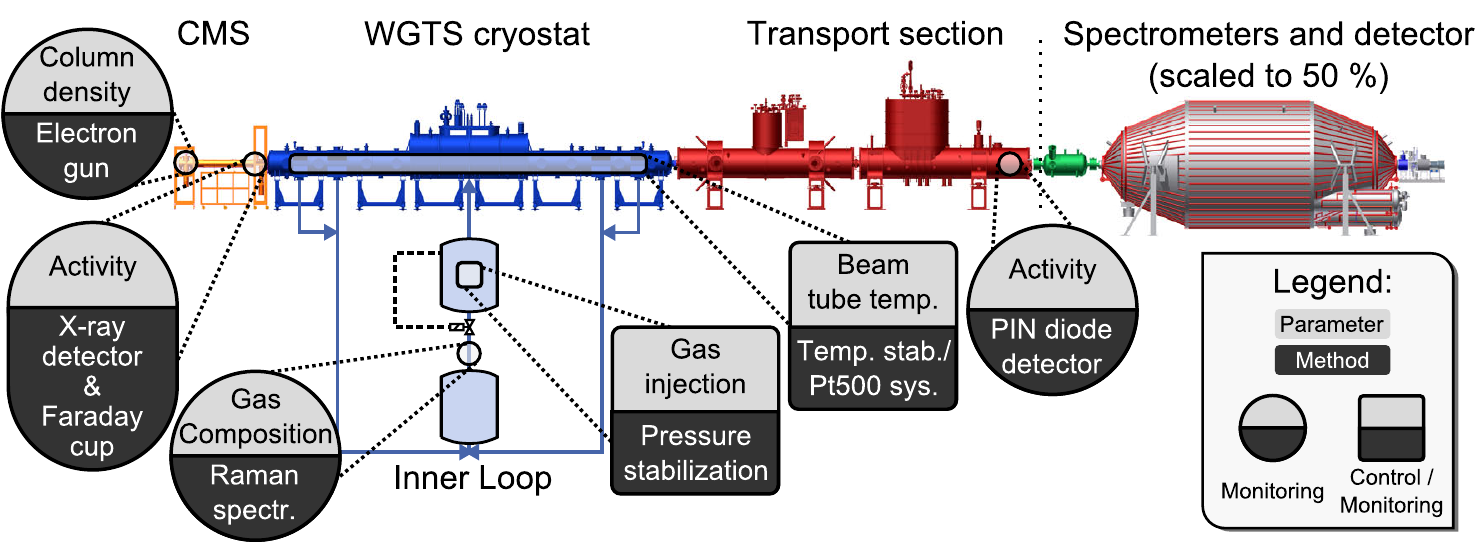}
	\caption{\textbf{Overview of the analytic systems for control and monitoring of the gaseous tritium source.} The detailed description of the systems is found in the main text. Note that both spectrometers are scaled to $50\,\%$ of their size.}
	\label{fig:katrin-monitoring-systems}
\end{figure}

The different control and monitoring systems summarized in figure~\ref{fig:katrin-monitoring-systems} are:
\begin{itemize}
	\item[-] \textbf{The Inner Loop system} provides stable gas injection and the stable gas composition together with the infrastructure of the Tritium Laboratory Karlsruhe (TLK) (see section~\ref{sec:loops}).
	\item[-] \textbf{A sensor system of PT500 temperature sensors} provides the monitoring of the spatial and temporal temperature distribution $T_{BT}(\phi,z)$ on the WGTS beam tube (see section~\ref{sec:BTTTemperatures}).
	\item[-] \textbf{The Laser Raman system (LARA)} is connected to the Inner Loop and enables the in-line monitoring of gas composition $\epsilon_T$ (see section \ref{sec:LARA}).
	\item[-] \textbf{Three activity detectors} are foreseen to measure the activity of the WGTS. The overall WGTS activity is proportional to the product of column density (\rhod) and isotopic purity $(\epsilon_T)$ (see section~\ref{sec:activity}). 
	\item[-] \textbf{An electron gun} embedded into the Calibration and Monitoring System (CMS) of the KATRIN experiment is used to measure the scattering losses of the electrons in the WGTS (see section~\ref{sec:e-gun}).
\end{itemize}
It should be noted that most of the systems now have been set up, and we have proven their suitability for the various control and monitoring tasks. The few remaining systems are already beyond the prototype phase and the results thus far suggest that they could be implemented into KATRIN in the near future. 
Below we introduce the various control and monitoring systems, and discuss their current status.


\subsection{The inner tritium loop}
\label{sec:loops}
From the point of view of tritium circulation, achieving a stable column density requires (i) the stable injection of tritium in the middle of the WGTS beam tube, and (ii) stable pumping on both ends.

While conceptually this sounds very simple, in reality it requires a rather complex assembly. Overall, the actual working principle very much resembles Ohm's law (for gases):
$Q=\Delta p \cdot c$. A constant pressure gradient $\Delta p$ and a constant conductance $c$ (which will be achieved by a constant temperature and length of the connection pipe) will result in a constant gas flow $Q$ into the WGTS. 

\begin{figure}[tb]%
\includegraphics[width=\columnwidth]{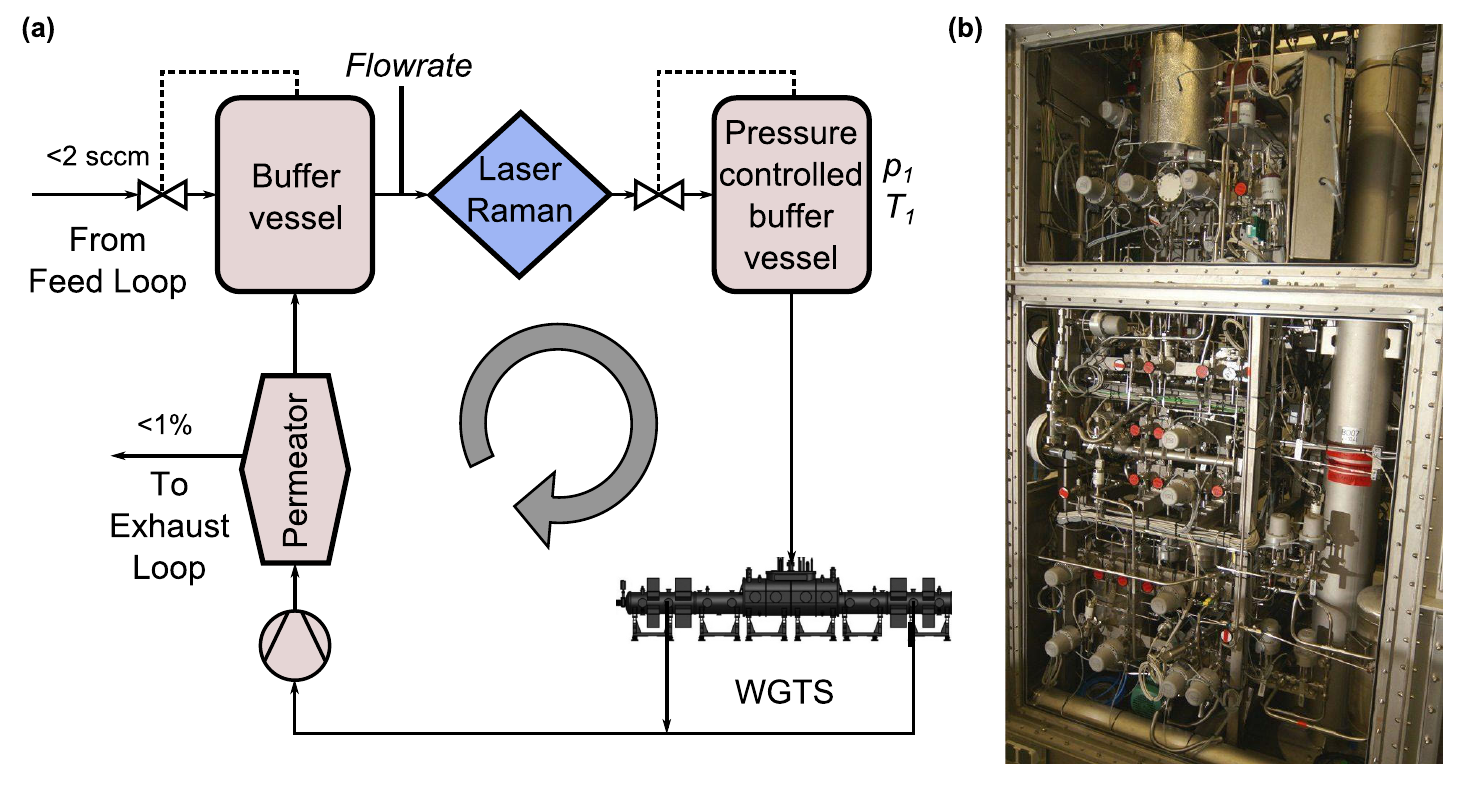}%
\caption{\textsl{(a)} Simplified flow scheme and \textsl{(b)} picture of complete setup of the tritium circulation loop.
\label{fig:Loop_Flussdiagramm}}
\end{figure}

To fulfil the first requirement, a highly precise ``flow controller`` had to be built at the Tritium Laboratory Karlsruhe; this is the so-called Inner Loop (figure~\ref{fig:Loop_Flussdiagramm}, (b)). It consists of
45 sensors for pressure, temperature and ionization, 70 automatic and manual valves, 2 flow controllers, 1 flow meter, 9 vacuum pumps, 6 buffer vessels, 2 palladium membrane filters, $\approx70$ custom made pipes, $\approx20$ metal bellows and a total of $>400$ metallic seals.

The second requirement, i.e. a stable pumping rate, will be achieved using turbo molecular pumps (Oerlikon Leybold MAG W 2800) on both ends of the WGTS. Due to a constant rotation frequency (on a $10^{-4}$ level), a constant pumping speed and consequently a constant exit pressure will be maintained.

After setting up the flow controller, leak tests of every single component and an integral leak test were performed. The measured integral leakage rate of $7\cdot 10^{-9} \,{\rm mbar}\cdot {\rm l/s}$ conforms with the regulations for primary systems of the Tritium Laboratory. 

Since the WGTS is not yet available for test measurements, the connection had to be bypassed by a tube with similar conductance as the later combination of WGTS and connection line.

\begin{figure}[tb]%
\centering
\includegraphics[width=\columnwidth]{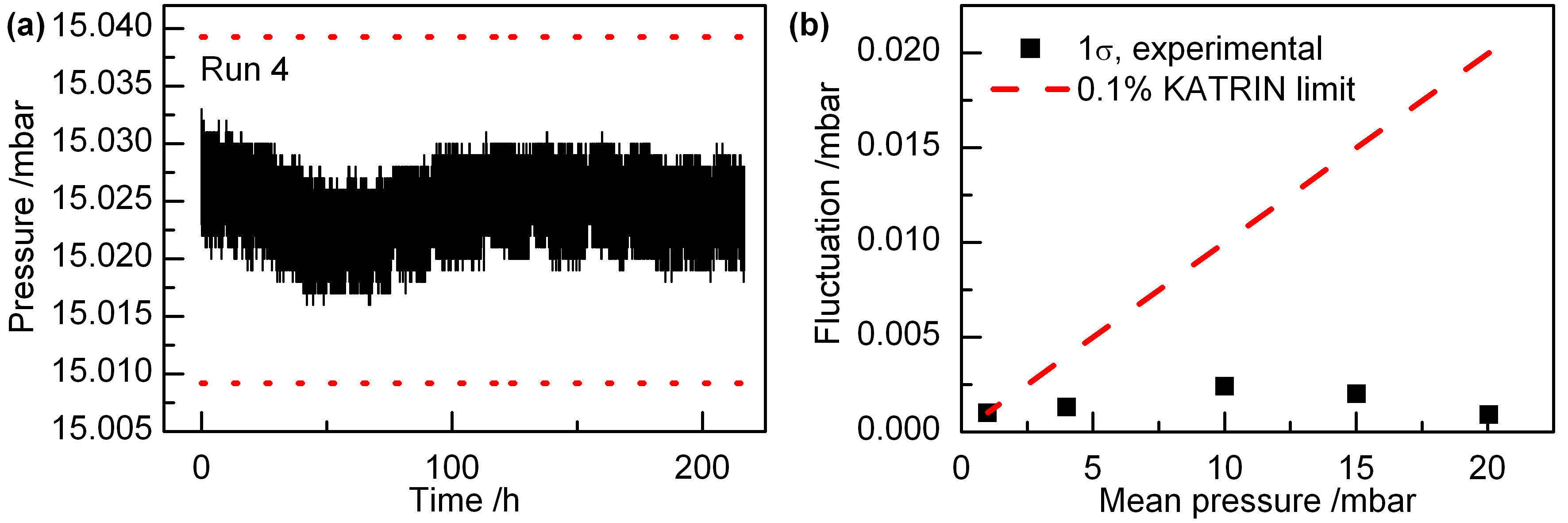}%
\caption{\textsl{(a)} Results of an individual pressure-control run. The pressure in the WGTS buffer vessel is shown for several hours of deuterium circulation and a pressure set-point of $15\,{\rm mbar}$. The dashed line represents the $1\sigma$ requirement for KATRIN operation. \textsl{(b)} Measured pressure fluctuation as a function of the mean pressure during nitrogen circulation, as measured for different set-points. }%
\label{fig:Loop_Diagramme}%
\end{figure}

An example for the measured pressure trend during pressure-stabilized circulation of deuterium is shown in figure~\ref{fig:Loop_Diagramme}. 
Here, the KATRIN limits are indicated by the dotted lines; clearly, the pressure stabilization falls well within these limits. Note that the particular run shown here may be seen as the worst-case. The apparent dip during the early times of the run did not occur in any other of the runs; it could well have been the result of a variation in the room temperature.
From the trace in figure~\ref{fig:Loop_Diagramme}, (a), one extracts an average pressure $\langle p_1 \rangle = 15.024 \pm 0.002$~mbar inside the pressure-controlled buffer vessel, for a set-point of 15~mbar.

The corresponding gas flow is stable within 120.4$\pm$0.5~sccm, with a temperature fluctuation of the buffer vessel of $\pm$0.125~K at 318~K.
Fourier analysis of the pressure trends shows no indication for periodic short term (seconds/hours) or long term (days/weeks) fluctuations. 

For calibration (see e.g. section~\ref{sec:e-gun}) and verification reasons, the pressure stabilization is designed to operate between $p_1 \approx1\,$mbar and $p_1 \approx20\,$mbar. 
A series of pressure stability tests at different pressures has been made with N$_2$ filling. The results are shown in figure~\ref{fig:Loop_Diagramme} (b). The observed fluctuations are well within the KATRIN requirement for all pressure set-points.
In particular, for the standard KATRIN conditions with a pressure set-point of $p_1 \approx15$~mbar and only negligible temperature fluctuations on the connection line to the WGTS, the Inner Loop performs nearly an order of magnitude better than the $0.1\%$ limit required by KATRIN, for both deuterium and N$_2$.

To examine the effect of a temperature variation of the WGTS connection line, the WGTS bypass is equipped with an electrical heater to simulate temperature changes resulting in a different conductance. The temperature $T_{{\rm bypass}}$ was varied in four steps over an interval of $\approx$ 20~K within a few hours. This exceeds the expected fluctuations under KATRIN conditions by three orders of magnitude. 
Such extreme variations of up to 20~K do not affect the pressure to such an extent that the KATRIN $0.1\%$ limit would be surpassed. For a variation $\Delta T_{{\rm bypass}}$ of $\approx2$~K (still $\approx 100$ times the KATRIN specification), no visible influence on the pressure stabilization has been observed.
Therefore, it can be concluded from these test measurements that the Inner Loop fulfils and even exceeds the requirements for KATRIN.

\subsection{Beam tube temperature}
\label{sec:BTTTemperatures}

In both the WGTS demonstrator setup and the final WGTS assembly described in section~\ref{sec:physicalwgts}, the monitoring of the beam tube temperature is performed by 24 metallic resistance temperature sensors (Pt500) and 24 vapour pressure sensors. The Pt500 sensors are used to monitor the temperature continuously. Their resistance is measured by digital multimeters with a resolution of $\Delta R = 3\cdot10^{-5}\,\Omega$. A common calibration curve, measured for the original sensors of the WGTS, is used to convert the resistance to actual temperatures with a theoretical temperature resolution of $\Delta T = 3\cdot10^{-5}\,{\rm K}$. Several contributions to the uncertainty have been evaluated in~\cite{grohmann_precise_measurement}; the results suggest a combined calibration uncertainty of $125\,{\rm mK}$ at $T=30\,{\rm K}$; the main contribution seems due to the magnetic field dependence of the Pt500 temperature measurements. This uncertainty is insufficient to reliably monitor the beam tube temperature with requirements of 30~mK on stability and homogeneity.
Therefore, each Pt500 sensor is calibrated in situ by an adjacent vapour pressure sensor. A vapour pressure sensor is a small volume filled with LNe. The temperature is determined by measuring the saturation pressure of neon with a transducer outside of the cryostat, connected to the sensor by a small capillary. This measurement is insensitive to magnetic fields and has a measurement uncertainty of only 4~mK. However, continuous measurement are not possible since the neon filling of the sensors can neither be guaranteed nor verified during the expected KATRIN run periods of up to 60 days. Instead, the very accurate vapour pressure sensors can be used to calibrate the Pt500 sensors before and after every KATRIN tritium run to reduce the uncertainty of each individual sensor to that of the LNe pressure sensors, i.e. 4~mK. This is sufficient to monitor the beam tube temperature.

As mentioned in section~\ref{sec:physicalwgts}, this approach was tested in 2011 with the WGTS demonstrator test setup. 
Data on the stability of the primary LHe supply and thermosiphon (see figure~\ref{fig:thermosiphon}) have already been analysed. 

\begin{figure}[tb]%
\includegraphics[width=\columnwidth]{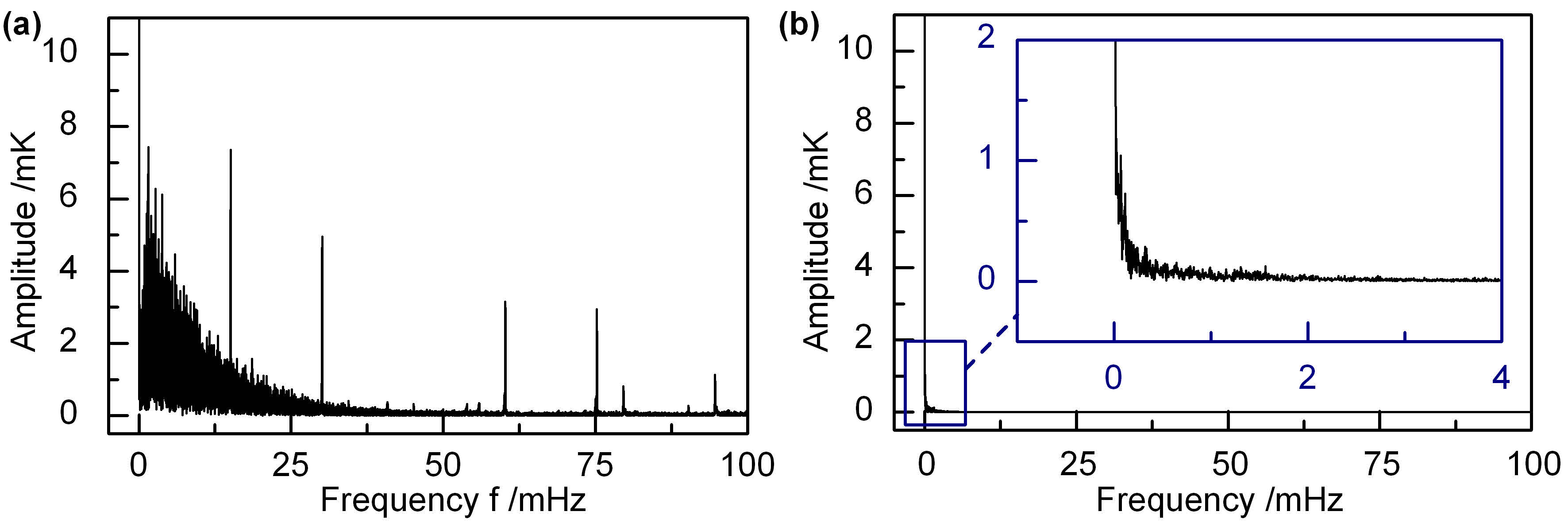}
\caption{\textsl{(a)} Discrete Fourier transform of the temperature of the original cryogenic helium used to cool the neon system for a 2 day measurement period. \textsl{(b)} Corresponding Fourier transform of the signal of one of the temperature sensors on the beam tube for the same period. }%
\label{fig:demoFTs}%
\end{figure}

In particular, the shielding of the two-phase neon system from the mass flow and temperature fluctuations of the primary helium circuit, which are of the order of 300~mK~\cite{grohmann_stability_analyses}, works as expected. This is evident from the discrete Fourier transforms of both the primary helium circuit and beam tube temperature shown in figure~\ref{fig:demoFTs}: While the helium circuit exhibits significant fluctuations and characteristic frequencies, only uncritical low-frequency fluctuations on the time scale of days are transferred from the helium circuit through the condenser to the two-phase neon system; all critical frequencies with $f>0.3\, {\rm  mHz}$ are reduced by at least two orders of magnitude. This indicates that the temperature stability is well within the design specifications for KATRIN.
However, the full analysis of the data taken at the WGTS demonstrator is still ongoing and will be reported in a future publication~\cite{grohmann_demonstrator}.

\subsection{Gas composition monitoring using Raman spectroscopy}
\label{sec:LARA}
As stated in section~\ref{sec:IsotopologicalComposition}, the isotopic purity $\epsilon_T$ of the inlet gas of the WGTS has to be monitored continuously with a precision $< 10^{-3}$. The integration time for an individual measurement should be as short as possible to be compatible with the measurement intervals of KATRIN. 
Typical methods for quantitative analysis of tritium, e.g. gas chromatography~\cite{GC} and liquid scintillation counting~\cite{LSC}, are not suitable for non-stop operation at the KATRIN experiment because of the duration of the measurement procedure (typically $\gg$~250~s), the need of sample extraction and the continuous waste production. In contrast, Laser Raman spectroscopy is a non-contact and in-line method which is based on the inelastic Raman scattering~\cite{Long} of light from gas molecules and continuous monitoring in measurement intervals of $<250$~s is feasible. Raman spectroscopy systems are commercially available and widely used for gas analysis. But when combining the requirements of operation with tritium and $<10^{-3}$ statistical uncertainty, off-the-shelf commercial products are not suitable~\cite{LARA-Schloesser}.

Although, in principle, no other gas species than hydrogen isotopologues are expected to be injected into the source tube, Raman spectroscopy is also sensitive to any other molecular species like nitrogen and tritiated methane species (CT$_4$, CDT$_3$, etc.)~\cite{LARA-fischer}; the former is thought to originate from incomplete evacuation of the loop prior to the filling with tritium, while the latter can be attributed to exchange reactions with carbon liberated from the steel vessel and tubes.

\begin{figure}
	\includegraphics[width=0.49\columnwidth]{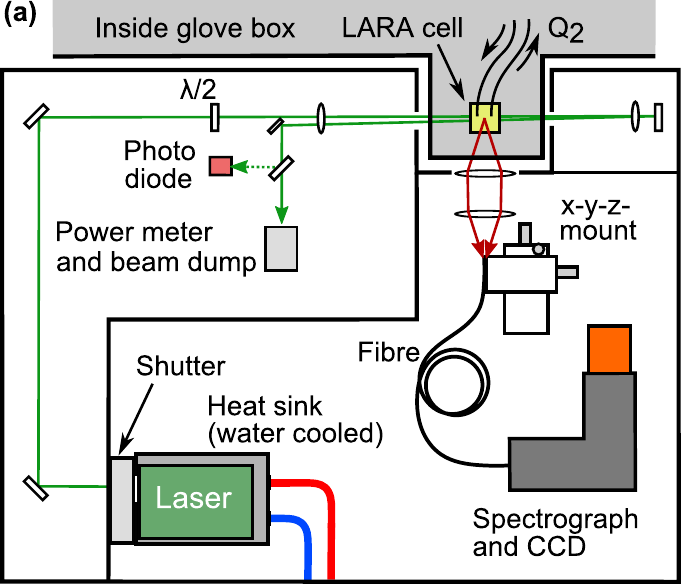}
	\includegraphics[width=0.49\columnwidth]{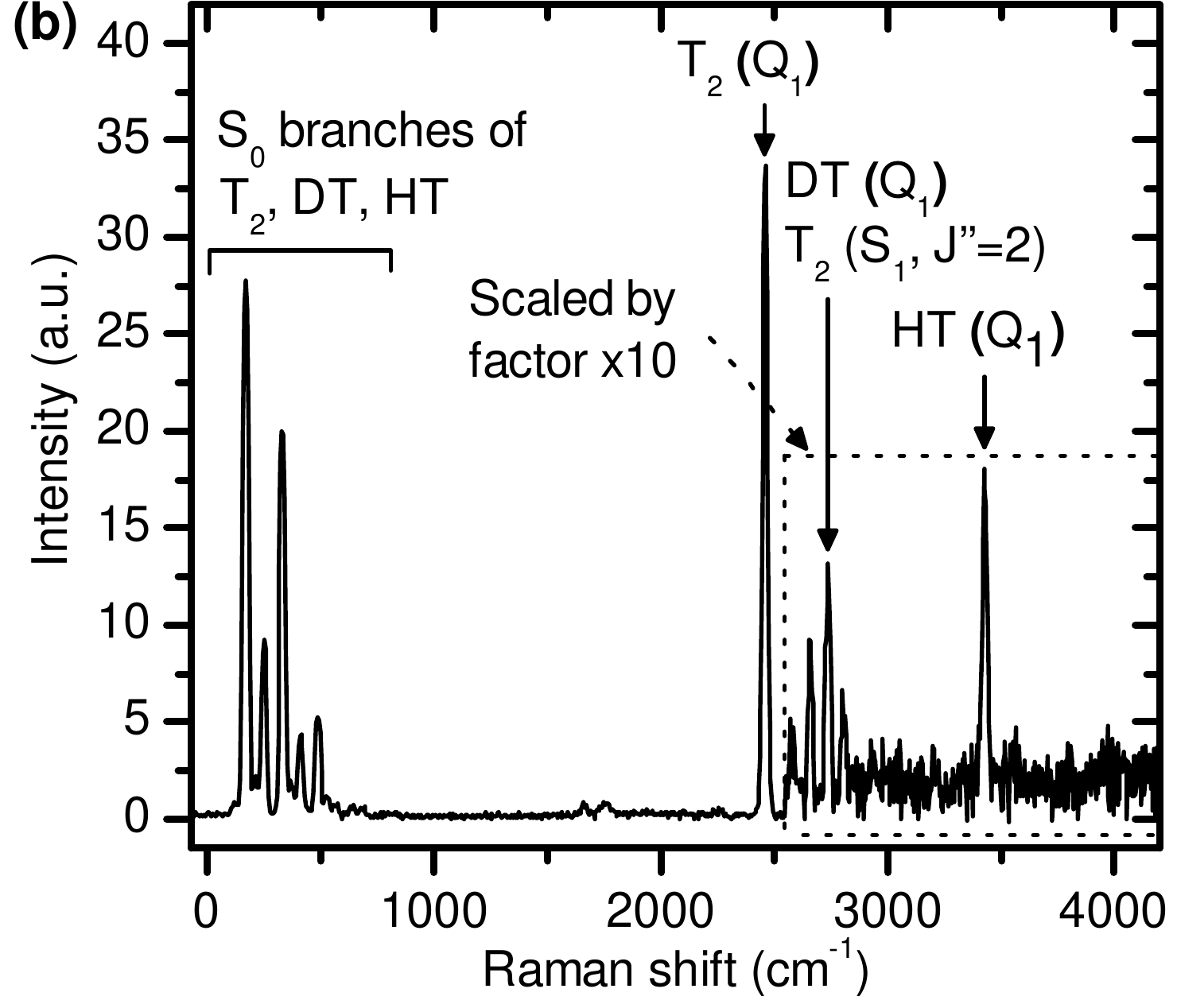}
	\caption{\textsl{(a)} Simplified schematic drawing of the LARA setup (for details see main text). The polarization angle of the laser beam is controlled by a half-wave plate ($\lambda$/2). The photo diode and power meter are used for monitoring of the laser stability. The LARA cell is located inside the glove box of the Inner Loop while the optical setup is on the outside to prevent contamination. \textsl{(b)} Representative Raman spectrum of a gas mixture with p$_\mathrm{tot}$ = 217~mbar and $\epsilon_T >90\%$. The Raman branches are labeled according to~\cite{Long}. T$_2$ dominates the mixture but traces of DT and HT are visible.}
	\label{fig:LARA-setup}
\end{figure}

Over the past five years, a specialized Laser Raman spectroscopy unit (LARA) has evolved and was thoroughly tested~\cite{LARA-Lewis, LARA-Sturm}. It consists of a measurement cell (LARA cell), which is located between the buffer vessels of the Inner Loop (see figure~\ref{fig:Loop_Flussdiagramm}) and a laser setup for gas analysis (see figure~\ref{fig:LARA-setup}, (a)). The gas pressure inside the LARA cell is typically in the range of 150--200~mbar during operation of the Inner Loop. A DPSS laser, operating at 532~nm, is used as excitation source. The Raman scattered light is collected under an angle of 90$^\circ$ by optical lenses and guided by an optical fibre to a spectrometer (Princeton Instruments, HTS) and CCD detector (Princeton Instruments, PIXIS:2KB) for spectral analysis. The Raman shifts and intensities of the vibrational Q$_1$ branches of the hydrogen isotopologues are used for qualitative and quantitative analysis, respectively (see figure~\ref{fig:LARA-setup}, (b)). A custom-written analysis package has been developed, that incorporates the automated correction of high intensity CCD pixels due to cosmic rays, background subtraction and peak fitting~\cite{LARA-paper3}.

Although the present system is still based on the originally proposed concept~\cite{TDR}, recently three major improvements have been implemented which aided in maximizing the signal-to-noise ratio:
\begin{itemize}
	\item[-] A temperature controlled high-stability laser (Laser Quantum, Finesse 5~W, 532~nm), and thermally stable mounts for the optical components are used to reduce fluctuations of laser power and beam alignment that would impose intensity fluctuations in the Raman spectra. 
	\item[-] The laser beam is passed through the LARA cell twice to nearly double the effective laser power in the cell with respect to the standard, single-pass setup. 
	\item[-] Groups of CCD pixels are simultaneously read-out (``on-chip binning``) to minimize noise from the analogue-to-digital conversion in the CCD~\cite{LARA-paper3}. 
\end{itemize}
By 2011, the LARA system was tested in the tritium test circulation loop, LOOPINO. Here it had already reached $\Deltaprec(\epsilon_{\rm{T_2}})/\epsilon_{\rm{T_2}}=0.1\%$ for T$_2$ monitoring within 250~s acquisition time under conditions similar to KATRIN operation ($\sim$200 mbar absolute pressure, tritium purity $\epsilon_T \sim 0.95$) before the aforementioned improvements were implemented~\cite{LARA-fischer}. At this time, a level of detection of 0.012~mbar partial-pressure equivalent was reached~\cite{LARA-Schloesser}. After implementation of the improvements, a first test with a tritium mixture ($p(\rm{ T_2})~\approx~7.4$~mbar, $p({\rm{HT}})~\approx~1.6$~mbar, $p({\rm{DT}})~\approx~0.7$~mbar) was performed. For this, a statistical uncertainty of $\Deltaprec(\epsilon_{\rm{T_2}})/\epsilon_{\rm{T_2}}~=~0.3\%$ was achieved for T$_2$ monitoring within 250~s acquisition time. Extrapolating this performance to the KATRIN operating conditions, the LARA system can reach $\Deltaprec(\epsilon_{\rm{T_2}})/\epsilon_{\rm{T_2}}=0.1\%$ within $60\,{\rm s}$ acquisition time. 

In contrast to the other monitoring tools of the WGTS, not only the precision but also the trueness of LARA measurements is of interest, due to the influence of the different final state distribution of the individual isotopologues on the $\beta$-spectrum (see section~\ref{sec:IsotopologicalComposition}). 

To address the problem of trueness in the observed Raman signal amplitudes, a calibration method with non-trititated reference gas mixtures has been developed to determine the response functions for H$_2$, HD and D$_2$, which relate the Raman line intensities to the molar fraction of the isotopologues in the gas mixture. The response functions, which are different for the individual isotopologues, include quantum mechanical transition matrix elements, the $\tilde{\nu}^4$ wave number dependence of Raman scattering and the spectral sensitivity of the light collection system of LARA. In this study, an apparent variation between the individual response functions by up to 10\% has been observed.

In an alternative approach, ``synthetic`` calibration spectra are generated against which the measured spectra are compared. For this, the transition matrix elements of the hydrogen isotopologues are calculated using \emph{ab initio} formalisms~\cite{LeRoy}. Unfortunately, the relevant matrix elements were not yet experimentally verified for all hydrogen isotopologues; hence, any slight errors therein would potentially propagate into the LARA measurement of the gas composition. Therefore, a precise experimental check of the transition matrix elements of all hydrogen isotopologues has been performed and results will be reported shortly~\cite{LARA-paper1}. 

Another calibration issue in determining the trueness in the Raman peak amplitudes is the spectral efficiency of the light collection system; for this a complete characterization of the LARA setup is required, which will rely on a well-defined calibration light source (presently under development). The aim is to further improve the accuracy of the LARA calibration (currently $\sim10\%$). However, should the understanding of the final states improve, and thus the uncertainty estimate reduce, then this would immediately impact on the LARA accuracy requirements. Most likely they would become more stringent; however, the current improved implementation of the LARA setup would certainly be capable to meet such a demand.

As a final note, we would like to mention that during operation with concentrated tritium gas mixtures, damage of the tritium facing anti-reflection coatings of the LARA cell windows was observed~\cite{LARA-fischer}. A test experiment is currently ongoing to search for tritium resistant coatings and to investigate the potential loss of optical transparency of the windows under $\beta$-radiation. For final testing, a long-term measurement campaign of the LARA setup with a circulating tritium gas mixture ($\epsilon_T \geq 0.95$) is planned for 2012.

\subsection{Activity measurements}
\label{sec:activity}
The activity monitoring systems presented in this section have two important goals: (i) They provide information about fluctuations of the WGTS activity itself with a time-scale of minutes; these would cause systematic effects in the neutrino mass analysis if not taken into account. (ii) Together with the measured tritium purity $\epsilon_T$, equation~\eref{eq:signal} can be used (in the form $\rhod = S_i/(C_i\cdot \epsilon_T)$) to monitor fluctuations of the column density ${\cal N}$ with 0.1 \% precision. Here, the index $i$ reflects the fact that the proportionality constant $C_i$ and thus the signal rate $S_i$ is different for each activity monitoring system. The combined monitoring of both activity and gas composition thus allows one to disentangle whether observed activity fluctuations are due to changes of the gas composition or changes in the column densitity \rhod (which have a different effect on the KATRIN count rate).

Various concepts to measure the source activity have been studied. In all methods, a lower threshold for the $\beta$-electron detection is necessary to separate the signal from the overwhelming amount of low-energy secondary and shake-off electrons. Furthermore, the detection method must not disturb the electron transport from the source to the main spectrometer. 

In the following, we present three experimental concepts and devices which can be integrated into the KATRIN beam line. Two of these are located in the Control and Monitoring System (CMS), with either directly measuring the current in the rear wall incurred by the $\beta$-electrons (see section~\ref{sec:Faraday}), or with Beta-Induced X-ray spectroscopy(see section~\ref{sec:trirex}). The third device, the Forward Beam Monitor Detector (see section~\ref{sec:FBMD}), is located in the Cryogenic Pumping Section (CPS). Prototype tests demonstrated that with all three concepts the detection of electrons with a suitably low energy threshold is feasible.

The column density will also be measured periodically (every few hours) by an electron gun; as discussed in section~\ref{sec:e-gun}, this is an important calibration measurement by itself. It also allows to probe whether the constants $C_i$ of the activity monitoring systems are truly time independent (within the required precision of 0.1\%).

\subsubsection{The rear wall as a Faraday cup} 
\label{sec:Faraday}
To use the rear-wall as a Faraday cup to monitor the WGTS source activity, the WGTS $\beta$-electron emission is treated as a DC current, which is of the order $-10\,{\rm nA}$.  This is accompanied by low-energy plasma currents, which must be screened. For this a multi-layer thin-film coating is applied to the rear wall. The top layer is a gold film of about 50~nm thickness; while relatively thin, this gold layer is sufficient to intercept all low-energy plasma species. It is electrically isolated from the metallic substrate by a $150\,{\rm nm}$ ${\rm SiO_2}$ insulator of $10^6-10^7\,\Omega$. This dual-layer thin-film structure is sufficient to meet the aforementioned threshold conditions: electrons above $\sim$5~keV reach the substrate, while lower-energy electrons are stopped in the gold layer.  A high-precision ammeter can then measure the substrate current to $0.1\%$ precision in $0.1\,{\rm s}$. The radiation dose to the ${\rm SiO_2}$ layer is of the order of 20 MGy/y. In fact at such doses performance degradations for ${\rm SiO_2}$ capacitors are not expected. Such exposures will be tested in advance also to check the stability of the space charge accumulation.  

\begin{figure}
	\centering
		\includegraphics[width=1.00\textwidth]{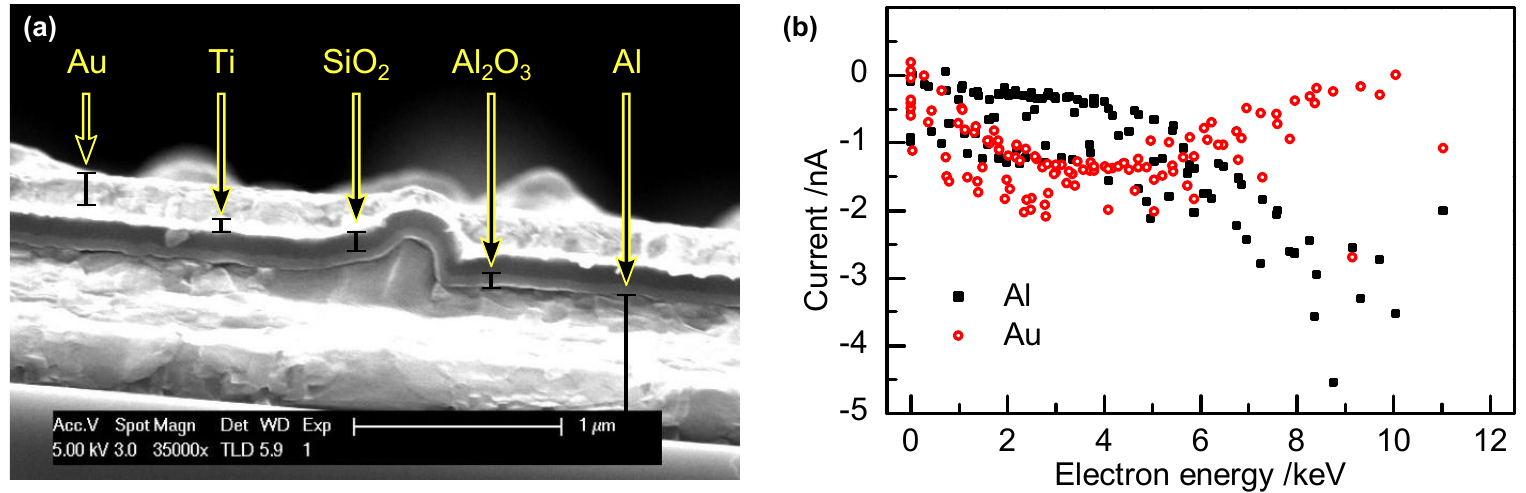}
	\caption{(a) Cross-section through a multi-layer rear-wall sample. (b) Measurement of the currents from the aluminium and gold layers in a different sample, as a function of the electron beam energy. This measurement displays the ability of a multilayer rear wall to separate low-energy from high-energy electrons.}
	\label{fig:rear-wall-pic-n-measurement}
\end{figure}

A mix of atomic-layer deposition (ALD) and plasma-enhanced chemical vapour deposition (PECVD) on an Al test substrate, is used to produce the Au/SiO$_2$ thin-film structures described above. For aluminium test substrates we achieved intact insulator areas as large as $30\,{\rm cm^2}$. These test samples were exposed to an electron beam of a few nA, and the current impinging onto the Au-layer and through the Al-substrate were measured simultaneously, as a function of electron energy. One observes that low-energy beams stop in the gold, high-energy beams penetrate down in the substrate, with the crossover energy (equal currents in both layers) at $\sim6\,{\rm keV}$.
The cross section of the layer structure for a selected sample is shown in figure \ref{fig:rear-wall-pic-n-measurement}, together with preliminary current measurements at different electron energies. The displayed data verify that this multilayer design fulfils the anticipated, basic functionality. Further systematic studies are under way (i) to measure the uniformity of the gold properties; (ii)to expose the insulator to high radiation doses and  (iii) to produce a sample of full size ($200\,{\rm cm^2}$) on the desired beryllium substrate.

If the on-going ${\rm R\&D}$ is successful and if it is possible in such a current measurement to discriminate the arriving primary $\beta$-electrons from the low-energy secondary electrons, then this method will allow to monitor column density changes over the whole source cross section within measurement times of just a few seconds.

\subsubsection{X-ray detection for measuring the source activity}
\label{sec:trirex}
Most of the electrons from tritium $\beta$-decay in the WGTS ($\approx 10^{11}$ per second) are magnetically guided to the rear wall (including the electrons originally flying towards the spectrometer section, as most of them are reflected by the retarding potential and eventually hit the rear wall). The absorption of $\beta$-decay electrons in the gold coating of the rear wall then generates X-rays (see section~\ref{sec:physicalwgts}), which can be detected using Beta-Induced X-Ray Spectroscopy (BIXS). 

Provided that the rear wall substrate is sufficiently transparent for X-rays produced by bremsstrahlung or fluorescence, a detection system for activity monitoring of the WGTS can be placed behind the rear wall.

A proof of principle experiment (Tritium Rear wall eXperiment - TriReX) was built at TLK to demonstrate the feasibility of monitoring the activity of gaseous tritium under conditions similar to the WGTS. 

\begin{figure}[tb]%
\centering
\includegraphics[width=\columnwidth]{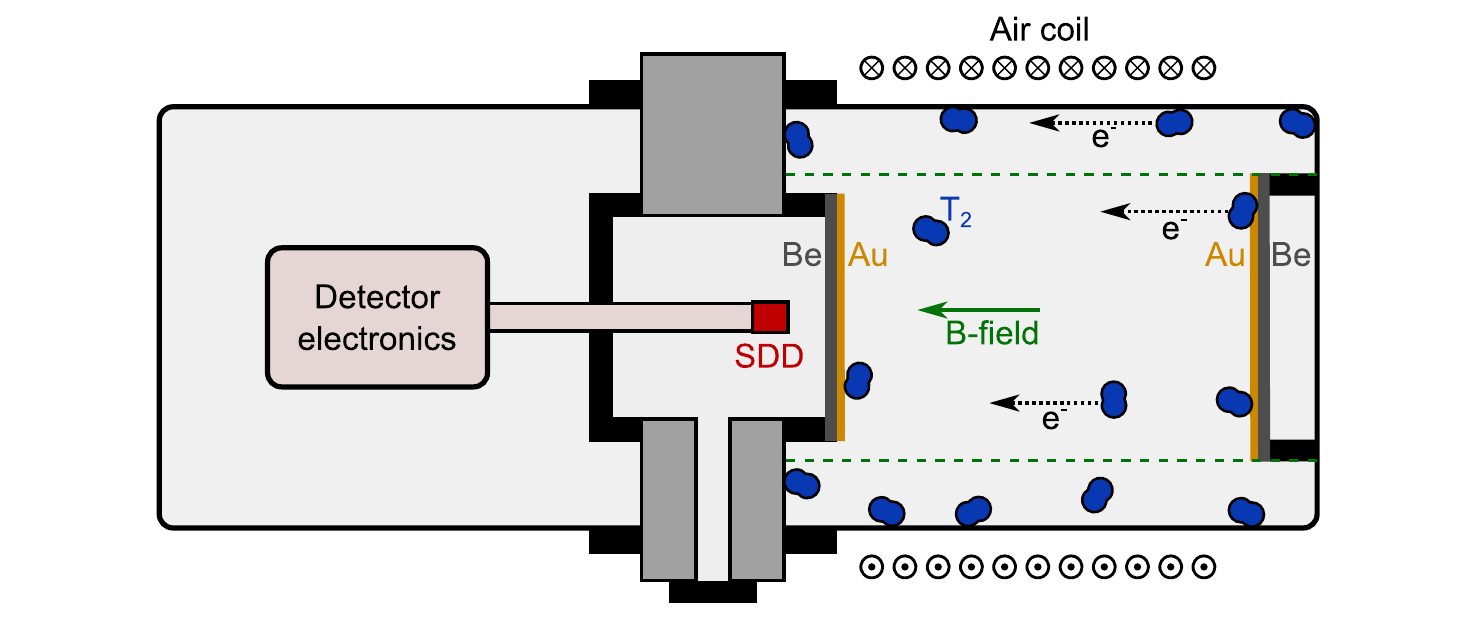}%
\caption{Schematic of the TriReX setup.}
\label{fig:TriReXsetup}%
\end{figure}

A simplified setup of TriReX is shown in figure~\ref{fig:TriReXsetup}. Conceptually, it is divided into three separate volumes.  The first (right) recipient volume contains the (H,D,T)-gas mixture. An axial magnetic field up to 120~mT provided by a water-cooled air coil guides the $\beta$-decay electrons to the gold coated beryllium windows at both ends. The right beryllium window (of 50~mm diameter and $180~\mu$m thickness) serves as a reference surface for later adsorption measurements, while the left beryllium window (with 39~mm diameter and $200~\mu$m thickness) constitutes the equivalent of the KATRIN rear wall. In addition, the window prevents contamination of the silicon drift detector (SDD, KETEK AXAS-M), located in the second (central) volume, with tritium. The SDD (active area: 80 mm$^2$)  is oriented towards the beryllium window. The energy resolution of the system is less than $160\,$eV at $5.9\,$keV. The third (left) volume contains the detector electronics, including pre-amplification and temperature control.

\begin{figure}[tb]%
\centering
\includegraphics[width=\columnwidth]{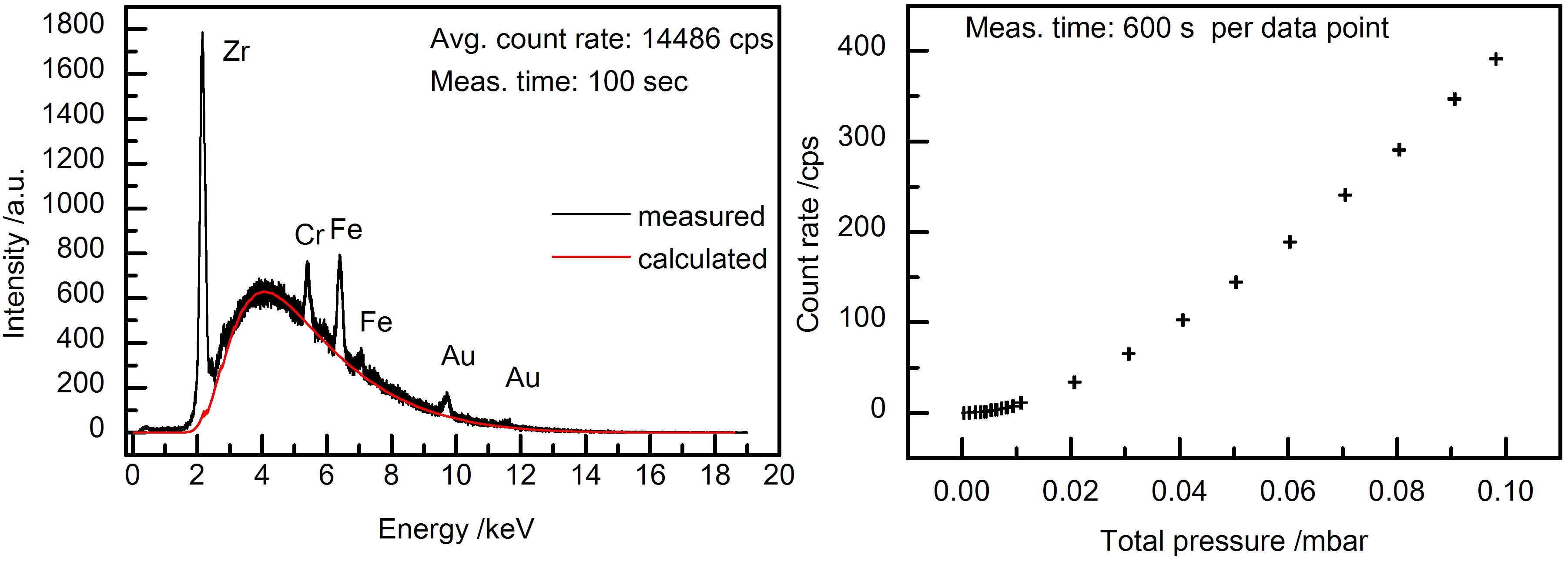}
\caption{\textsl{(a)} Bremsstrahlung spectrum and characteristic X-rays, measured with the TriRex setup, together with the calculated bremsstrahlung spectrum for a tritium partial pressure of $(0.8\pm0.1)\,$mbar and a total activity of $(1.4\pm0.1)\cdot10^{11}\,$Bq. The total intensity of the calculated spectrum was fitted to the measurement. \textsl{(b)} Corrected integral count rate vs. tritium partial pressure (error bars are included) taken with the TriReX setup. The corrected integral count rate is the integral count rate minus the memory effect signal $\dot{N} = (35.2\pm0.3)\,$cps after the evacuation of the system. The starting pressure of $0.098\,$mbar equates to $(8.3\pm0.8)\cdot10^9\,$Bq. The system was pumped down in successive steps to a final pressure of $0.001\,$mbar.}
\label{fig:TriReXspectrum}
\end{figure}

A spectrum taken with the TriReX setup is shown in figure~\ref{fig:TriReXspectrum}, (a). The continuous bremsstrahlung spectrum is superimposed by X-ray fluorescence lines. The X-ray line spectrum is dominated by a peak around 2~keV. This peak originates from the Zr-collimator which is factory mounted to the SDD. Additional peaks can be associated with Cr, Fe, Mn and Mo which are present as trace constituents in the stainless steel vacuum vessel.
 Two further peaks at higher energies are caused by the gold coatings of the beryllium windows. The characteristic X-rays provide the unique possibility of an in-situ energy calibration of the system. The cut-off below 2~keV is caused by absorption of the X-rays within the beryllium windows; the energy threshold of the detector had been set to $\leq 150\,$eV. The shape of the measured bremsstrahlung spectrum agrees well with calculations using the analytical model from~\cite{bib005} and the mass attenuation coefficients taken from~\cite{bib006}. 

With the observed count rate of $\dot{N} = 14486\,$cps, the necessary statistical precision of $\Deltaprec = 0.1\,\%$ is reached in $t\leq70\,$s measurement time within TriReX. For pressures lower than $\approx$10~mbar the integral count rate depends in principle linearly on the activity in the recipient~\cite{bib001}. Deviations from linearity as seen in figure~\ref{fig:TriReXspectrum} (b) are most likely caused by outgassing effects. The TriReX setup could not be baked out because of design limitations. Therefore outgassing from the walls during the pump-down causes an activity signal which is lower than expected because of the decreasing tritium purity. Also a tritium memory effect has been observed because of tritium adsorption on the walls. This memory effect causes an increasing background signal after each tritium measurement run.

The achievable count rate in the Control and Monitoring Section (CMS) for the WGTS setup depends on (i) the final geometry of the rear wall and the X-ray detector; (ii) the thickness and the layer composition of the rear wall; and (iii) the number of implemented X-ray detectors. However, the measured spectra here agree well with the calculated and simulated expectations. Therefore, a reliable extrapolation to the WGTS case is viable.

For the activity monitoring of the WGTS, the current design of the CMS includes two SDDs, with $100\,$mm$^2$ active area each. The thickness and the layer composition of the rear wall is still under investigation with R\&D efforts to coat a $\approx400\,\mu$m beryllium window with $120\,$nm of mono-crystalline gold. The expected count rate for this particular material and coating configuration would be $\approx6.6$ kcps, allowing the activity monitoring with $0.1\,\%$ precision within a sampling time of $\approx150\,$s.

Further investigations regarding the long term stability of such a system, backgrounds by tritium adsorption on gold and stainless steel and an absolute calibration of the system at the $\approx 1\%$ level will take place in the near future.

\subsubsection{Forward Beam Monitor Detector}
\label{sec:FBMD}		
\begin{figure}
	\centering
		\includegraphics[width=\columnwidth]{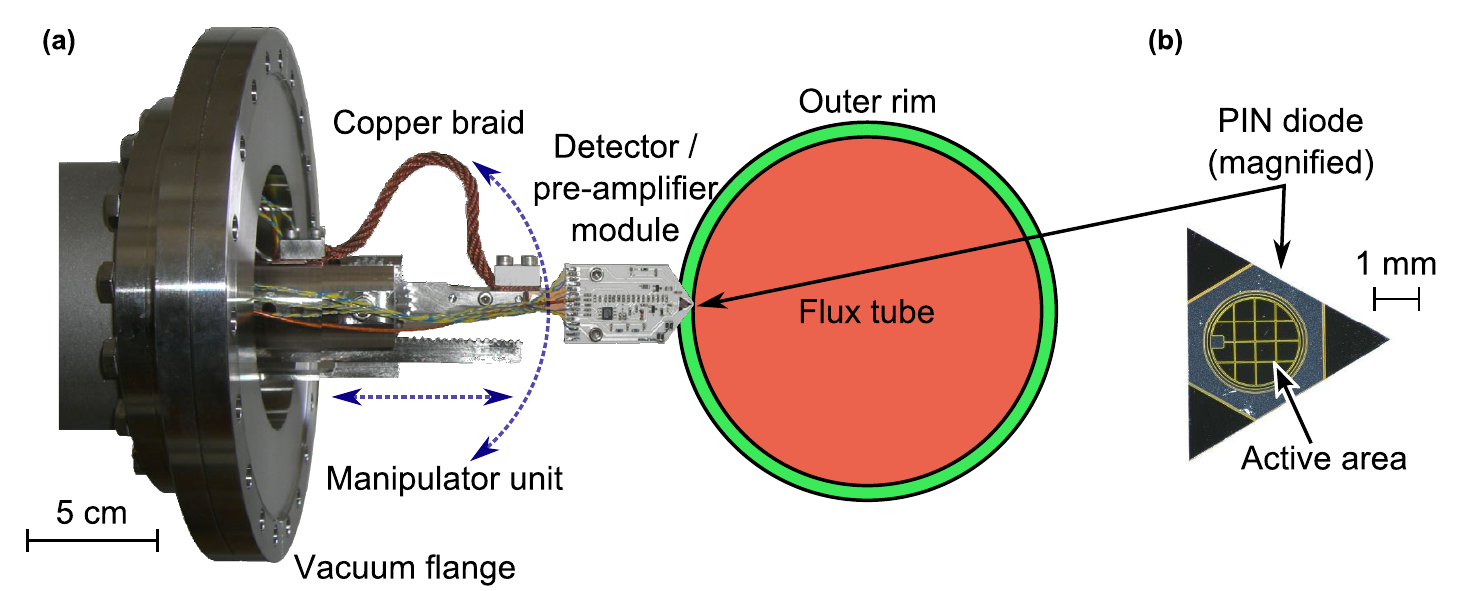}
		\caption{Setup of the Forward Beam Monitor Detector. \textsl{(a)} Manipulator with detector module in indicated flux tube (same scale). \textsl{(b)} PIN diode with an active area of $1.3\,{\rm mm}^2$.  }
	\label{fig:FBMD}
\end{figure}

Silicon radiation detectors, which directly detect the $\beta$-electrons in the flux tube, can also be used to monitor the source activity. Such a detector should not shadow any parts of the electron flux tube, which is used for the neutrino mass analysis, neither in the rear nor the forward directions. Therefore, care has to be taken where to position the detector. Because of these constraints, this component will be implemented in the so-called Forward Beam Monitor Detector (FBMD) configuration, which probes only the outermost rim of the flux tube. A picture of the setup is shown in figure \ref{fig:FBMD}. It is assumed that this localized activity measurement is nevertheless representative for the whole beam cross section. That this assumption holds may be verified during periodic calibration runs, since the detector concept also includes the option to move the detector element across the whole flux tube cross section. 
The FBMD will be mounted between the last two superconducting solenoids of the cryogenic pumping section (CPS), i.e. at the end of the transport section, where the Tritium partial pressure is reduced by 14 orders of magnitudes compared to the WGTS. This minimizes any background effects and contamination of the sensitive detector surface by tritium, and vice versa minimizes any impact from the detector on the Tritium gas.

The magnetic field $\vec{B}$ at this position is axially symmetric with $ \left|\vec{B} \right| \sim 1.6\,{\rm T}$ and can therefore be used to study the spatial homogeneity of the source profile with a position resolution of the manipulator of $50\,{\rm \mu m}$. For its standard monitoring role, the detector will be 61-67 mm away from the beam axis. 

The concept was tested by a proof-of-principle experiment using an off-the-shelf PIN diode (manufactured by Micron semiconductor). It comprises a circular active area of $1.3\,{\rm mm}^2$, a thickness of $300\,{\rm \mu m}$ and a dead layer of approximately $100\,{\rm nm}$.
The diode chip is mounted with epoxy on a ceramic substrate which also carries the charge-sensitive pre-amplifier as a hybrid module. The signal path is split into a spectroscopic AC mode and a DC mode which can be used in parallel. Both methods can be utilised to measure the actual activity of the WGTS. 

In the AC mode, the energy of each impinging electron is measured individually by digitalization of the pre-amplifier signal and subsequent pulse height analysis. 

In the DC mode, the continuous flow of decay electrons results in an increased DC offset of the integrating pre-amplifier, which scales with the activity. Of course, information on the energy of the $\beta$-electrons is lost in the DC mode, but sampling times are significantly reduced with respect to the AC mode; thus the activity can be measured with $0.1\%$ precision within just a few seconds. 

The reason for this is that in DC mode no ``pile-up`` effects are encountered, which are prevalent at higher count rates in AC mode. As a consequence, a larger detector area could be chosen for the DC mode detection, capturing a larger electron flux. The opposite would hold for the AC mode: count rates can be limited by making the detector area smaller. Currently, the sensitive detector area is a compromise between the optimal $0.1\,{\rm mm}^2$ for AC readout mode and $5\,{\rm mm}^2$ which would be the optimum for the DC readout mode. For the final design, it is proposed to use a custom designed detector chip with two active pixels, their respective sizes optimized for AC and DC readout mode.

For both modes, temperature stabilization of the detector is important.
In the spectroscopic (AC) mode, the energy resolution and thus the stability of detection threshold depends on the detector temperature.
In the DC mode, the diode leakage current which is superimposed on the signal produced by the impinging electrons, rises exponentially with the detector temperature. 
Therefore, a precise temperature control and stabilization of the detector element at temperatures of about $-30^{\circ}{\rm C}$ is necessary.
The cooling concept is based on a highly flexible copper braid with a cross section of $10\,{\rm mm}^2$ which thermally connects the movable detector holder to a copper heat sink. The heat sink is cooled by a cold gas supply which is based on liquid nitrogen in a dewar vessel evaporated with an electrical heater. With this approach, temperature stabilization to better than $\pm$\,0.1\,K has been achieved. 

The detector as well as its electronics and temperature control was mounted on a two dimensional UHV manipulator; the complete assembly was tested with an electron gun. The beam scanning capability of the manipulator was successfully shown. 
In the (spectroscopic) AC mode the monitoring of the electron flux proved to be reliable up to rates of about 50 kHz when pile-up becomes the limiting factor. Within the linear response regime, the electron energy resolution was measured to be $ \sigma ({\rm FWHM} ) = 1120 \pm 50\,{\rm eV} @ 17.5\,{\rm keV} $; this agrees with calculated noise sources like Fano and electronic noise. 

In the (integral) DC mode, count rates of up to 212 kHz have been measured with a systematic uncertainty of $1.9\%$. It should be noted that during long-term measurements drifts of the DC signal were observed; it was suspected that electrons hitting the corners of the detector surface create this effect. A detector optimized for this measurement mode, incorporating an additional shielding, should reduce or avoid these adverse effects.

It was possible to show that the FBMD system is capable to provide activity monitoring to a reasonable precision, however for the final KATRIN implementation some modifications are required. The dimensions at the KATRIN experiment dictate that a larger UHV manipulator, but of the same type and the same cooling concept, has to be used. The detector, especially its sensitive area, needs to be designed to accommodate the high $\beta$-intensity of $1.6\cdot10^6\,{\rm s}^{-1}{\rm mm}^{-2}$. Due to the low energy of the electrons, which is insufficient to create lattice defects, radiation damage is not an issue \cite{Wunstorf97}. But the detector sensitive area needs to be limited to reach an impinging rate of e.g. $50\,{\rm kHz}$ in the AC mode, meaning that within an integration time of $30\,{\rm s}$ an activity uncertainty of $0.1\%$ can be achieved. On the other hand, the DC detector can measure with a high rate like $5\,{\rm MHz}$ to reach a precision of $0.1\%$ within 1~s integration time, assuming that the systematic error remains constant.
 
\subsection{Other measurement methods for the column density}
\label{sec:e-gun} 
In this paper, the focus has been on tools for in-line monitoring and characterization of the WGTS, on time scales as close as possible to real-time. For completeness, we  briefly address a complementary, but off-line, method to infer $\rhod$, namely via the analysis of scattered electrons. The method is based on an approach which involves both the high-resolution main spectrometer and the focal plane detector. Consequently, the precision of the method is linked to the performance and precision of those two main KATRIN components. It has the potential to directly measure the energy losses caused by scattering, with a nominal accuracy of the probability for $n$-fold scattering $P_n$ of $\Delta_{{\rm acc}}(P_n)/P_n<0.1\%$, and it may be seen as one of the most crucial calibration measurements for the KATRIN experiment. 

Unfortunately, this monitoring method cannot be performed simultaneously with the neutrino mass measurements. Therefore, it will be used only in periodic tests rather than for continuous monitoring. Despite not being a near real-time method it is a vital cross-calibration check for $\rhod$\, and therefore it is worthwhile to briefly describe it here. 

An electron emitter (electron gun, e-gun) in the CMS based on the photo-electric effect produces quasi-mono-energetic electrons with stable intensity and with a well-defined angle $\theta$ (the angle between electron momentum and $\vec{B}$-field) and adjustable kinetic energy $E_0$~\cite{egun-muenster}. The parameter pair ($E_0, \theta$) can be varied over the full range of tritium-$ \beta $-electron values encountered in the WGTS, with a maximum angular spread of $ \sigma_{\theta} < 4^\circ$. The maximum kinetic energy is technically limited to 25~keV with a proposed maximum energy spread of $\sigma_{E_0} <0.2\, $eV in the low rate mode (R$\,<\,10^4\,$cps). 

The measurement principle relies on the correlation between the electron scattering probability and column density \rhod \,: changes in the WGTS column density can be invoked by the changes of probabilities for no, single or multiple scattering events. 
The scattering probabilities are solely dependent on the total cross section, the path length through the source and the column density. From the Poisson distribution, one finds that the probabilities for no-loss and single scattering of an electron entering the WGTS from the CMS amount to:
\begin{equation} \label{eq_P0}
	P_0(\theta)= e^{-\frac{\sigma\cdot \cal{N}}{\cos\theta}}
\qquad {\rm and} \qquad
	P_1(\theta)= e^{-\frac{\sigma\cdot \cal{N}}{\cos\theta}}\cdot \frac{\sigma\cdot \cal{N}}{\cos\theta}
\end{equation}

Due to the high energy resolution of the main spectrometer, the energy loss spectrum ${\rm d}N/{\rm d}E_{{\rm loss}}$ can be measured for a given condition of the electron gun $(E_0,\theta)$. Then, deconvolving the differential cross section ${\rm d}\sigma/{\rm d}E_{{\rm loss}}$ from the measured spectrum yields $P_n$~\cite{TDR}. The required differential energy loss function is either taken from models or derived from a special calibration measurement.
According to the formulae~\eref{eq_P0}, the column density \rhod \, can be monitored with the measured precision of $\Delta(P_1/P_0)$.
However, the method is not suitable as near-time monitoring since this would lead to a significant loss of measuring time for the neutrino mass campaign. Therefore, it is useful only for periodic examinations instead of continuous monitoring. Nevertheless, it can be regarded as complementary method to the other $\Delta\rhod$\, measurements described above.


\section{Conclusions}
\label{sec:conclusions}
In order to achieve the design sensitivity in the measurement of the neutrino mass, all main-task components in the KATRIN experiment need to operate with extreme stability over extended periods of time, and their behaviour needs to be well characterized. In particular, this is true for the WGTS which provides the essential tritium $\beta$-electrons from whose energy spectrum analysis the neutrino mass will be extracted. Here, we have presented the status of the control and monitoring system that is designed to ensure that the stability of the key WGTS parameters, namely the column density \rhod \ and tritium purity $\epsilon_T$ can be monitored and analysed as close to real-time as possible. For this task, theoretical modelling and experimental methodologies were developed; both were successfully combined confirming that the challenging requirements for the WGTS operation can be met.

Gas dynamics simulations were used to link the requirements for the precision and accuracy of  \rhod \, and $\epsilon_T$ to experimentally observable quantities like pressures and temperatures. Progress in this modelling includes (i) the incorporation of temperature inhomogeneities in the gas dynamics simulation, and (ii) a detailed investigation of the position-dependent velocity distribution in the WGTS.

The first strategy to monitor \rhod\, relies on monitoring the experimental parameters influencing \rhod, namely injection and outlet pressure as well as the beam tube temperature, with a precision determined by these simulations.
The pressure stability is realized by the Inner Loop system. The Inner Loop is operational and has reached and even exceeded the necessary relative stability of the injection pressure of $\Delta_{\textrm{prec}}(p)/p<10^{-3}$.
The performance of the beam tube cooling system and the corresponding temperature monitoring has been tested with an almost full scale demonstration setup of the WGTS using original components. Fourier analysis of the measured temperature fluctuations on the beam tube confirm the functionality of the cooling circuit design and in particular prove that temperature or mass flow fluctuations in the primary LHe circuit are not transferred to the cooling circuit of the beam tube. 

Complementary to the control and monitoring of the experimental parameters defining \rhod, another method has been developed to monitor fluctuations in \rhod\, close to real-time; it is based on measuring tritium purity and source activity simultaneously.
 
The tritium purity, which is also of interest in its own right, is hereby determined with a Laser Raman setup (LARA);  LARA has also exceeded its specification and has reached 0.3\% statistical uncertainty in T$_2$ monitoring of a low mole-fraction tritium sample. Based on this result, it is expected that the system reaches 0.1\% stat. uncertainty in $< 60$~s for KATRIN relevant gas compositions. The system was calibrated with non-tritated gas mixtures. For each isotopologue, specific response functions were determined which vary by up to $\sim$10\%. Further improvements of the accuracy are expected once quantum mechanical transition matrix elements are included whose accurate experimental verification has just been completed.

For the activity monitoring, three different approaches are pursued. Two methods of measuring the source activity can be implemented in the Control and Monitoring System (CMS) at the rear end of the KATRIN setup: the $\beta$-electron-current through the rear wall of the KATRIN experiment can be measured with a Faraday cup, and Beta Induced X-ray spectroscopy makes use of the X-rays produced by these $\beta$-electrons to monitor the source activity. The third option, the forward beam monitor detector FBMD monitors the edge of the flux-tube in forward direction towards the main spectrometer. All three options are on the R\&D way to demonstrate their suitability to measure with the required precision of $10^{-3}$ under KATRIN conditions.

Since most of the systems are now operational and perform beyond expectation, we are confident that the challenging control and monitoring requirements for the operation of the KATRIN WGTS will be met and even exceeded. As a consequence of this better-than-expected performance, it might even be possible to further reduce systematic uncertainties.


\ack
The authors wish to thank S. Mirz for the contributions to the assembly, commissioning and test of the LARA setup, and T. Bode and T. Wahl for the commissioning, test and operation of the WGTS Demonstrator, as well as the workshops and technicians of KIT for their support. 

This work has been supported by the Bundesministerium f\"ur Bildung und Forschung (BMBF) with project number 05A08VK2, the Deutsche
Forschungsgemeinschaft (DFG) via Transregio 27 �Neutrinos and beyond�, the Helmholtz Association (HGF) and the Helmholtz Alliance for Astroparticle Physics (HAP), and the Department of Energy, grant number DE-SC0004036,. We also would like to thank the Karlsruhe House of Young Scientists (KHYS) of KIT for their support.

\ded
We dedicate this paper to the memory of our Russian colleague, Academician
Vladimir M. Lobashev (1934-2011), the long-term head of the Troitsk neutrino mass experiment, 
who pioneered many scientific and technical issues discussed in this paper.

\end{document}